\documentclass[useAMS,usenatbib,referee]{mn2e}
\usepackage{graphicx}
\def\integral{{\it INTEGRAL}}
\def\suzaku{{\it Suzaku}}
\def\xmm{{\it XMM-Newton}}
\def\sax{{\it Beppo-SAX}}
\def\chandra{{\it Chandra}}

\usepackage{txfonts}

\title[The X-ray view of type 2  AGN]{Broadband study of hard X-ray selected absorbed AGN. }
\author[A. De Rosa et al.]{A. De Rosa$^1$\thanks{E-mail: alessandra.derosa@iasf-roma.inaf.it}, F. Panessa$^1$, L. Bassani$^2$, 
A. Bazzano$^1$, A. Bird$^3$, R. Landi$^2$, 
\newauthor A. Malizia$^2$, M. Molina$^4$, P. Ubertini$^1$ \\
$^1$Istituto di Astrofisica Spaziale e Fisica Cosmica (IASF-INAF), Via del Fosso del Cavaliere 100, 00133 Roma, Italy\\
$^2$Istituto di Astrofisica Spaziale e Fisica Cosmica (IASF-INAF), Via P. Gobetti 101, 40129 Bologna, Italy\\
$^3$School of Physics and Astronomy, University of Southampton, Southampton, SO17 1BJ, UK\\ 
$^4$Istituto di Astrofisica Spaziale e Fisica Cosmica (IASF-INAF), Via Bassini 15, 20122 Milano, Italy\\
}

\begin{document}

\date{Accepted 2011 November 8. Received 2011 October 11; in original form 2011 September 5}
\pagerange{\pageref{firstpage}--\pageref{lastpage}} \pubyear{2011}

\maketitle

\label{firstpage}
 \begin{abstract}
In this paper we report on the broadband X-ray properties of a complete sample of absorbed Seyfert galaxies hard X-ray selected with \integral.
Our sample is composed of 33 sources, of which 15 are newly discovered AGN  above 20 keV  (IGR  sources) while 18 are already known type 2 AGN (''known'').
For 17 sources (15 IGR + 2 ''known'' sources) we have performed a broadband analysis using both \xmm, and \integral-IBIS data. To have a full view of the complete sample we have then complemented the analysis of the 16 remaining sources with already existing broadband studies in the same range.
The high quality broadband spectra are well reproduced with an absorbed primary emission with a high energy cutoff and its scattered fraction below 2-3 keV, plus the Compton reflection features (Compton hump and Fe line emission).
This study permitted a very good characterization of the primary continuum and, in turn, of all the spectral features.

A high energy cut-off is found in 30\% of the sample, with an average value below 150 keV, suggesting that this feature has to be present in the X-ray spectra of obscured AGN.
The hard X-ray selection favours the detection of more obscured sources, with the log N$_{H}$ average value of 23.15 (standard deviation of 0.89).
The diagnostic plot N$_H$ $vs$ F$_{corr}$(2--10 keV)/F(20--100 keV) allowed the isolation of the Compton thick objects, and may represent a useful tool for future hard X-ray observations of newly discovered AGN.
We are unable to associate the reflection components (both continuum and Fe line) with the absorbing gas as a torus (as envisaged in the Unified Model),  a more complex  scenario being necessary. In the Compton thin sources, a fraction (but not all) of the Fe K line needs to be produced in a gas located closer to the black hole than the Compton thick torus, and this is possibly associated with the optical Broad Line Region, responsible also for the absorption.  We still need a Compton thick medium (not intercepting the line of sight) likely associated to a torus, which contributes to the Fe line intensity and produces the observed reflection continuum above 10 keV.
The so-called Iwasawa-Taniguchi effect can not be confirmed with our data.
Finally, the comparison with a sample of unobscured AGN shows that, type 1 and type 2 (once corrected for absorption) Seyfert  are characterized by the same nuclear/accretion properties (luminosity, bolometric luminosity, Eddington ratio), supporting the ''unified'' view. 

\end{abstract}

\begin{keywords}
galaxies: active -- galaxies: Seyfert -- X-rays: galaxies
\end{keywords}

%

\section{Introduction}

According to the widely accepted Unification Model (UM) for active galactic
nuclei ( Antonucci 1993, Urry \& Padovani 1995), type 2 and type 1 AGN are believed to share the same central region and emission mechanisms with the only difference being due to orientation.
In Seyfert 2, only optical narrow emission lines are visible, while in Seyfert 1 both narrow and broad emission lines are detected (Antonucci \& Miller 1995).
Type 2 AGN are seen edge-on, across a large amount of obscuring gas that prevents us from directly observing the central core and the broad line regions (BLR). This scenario is strongly supported by spectropolarimetric
observations of hidden broad-line regions (HBLRs) in several type 2 AGN, and by the  
 X-ray observations, demonstrating that Seyfert 2 galaxies usually have absorption columns greatly exceeding the Galactic ones (Cappi et al. 2006). Despite this, and in view of  several pieces of evidence coming from recent observations of different classes of AGN, the UM is still being debated.
In particular, the association of the Compton thick (a cold gas with N$_{H}\gtrsim \sigma_{T}^{-1} \backsim$ 10$^{24}$ cm$^{-2}$) obscuring medium with a  uniform torus at intermediate scale between the  BLR and the narrow line regions (NLR), has been questioned by the presence of column density variability on timescales of hours, requiring an X-ray absorber no larger than the BLR. Such changes have been detected in the brightest absorbed Seyfert galaxies, NGC 1365, NGC 4151 and NGC 4388 (Risaliti et al. 2007, Puccetti et al. 2007, Elvis et al. 2004). 
It has also been  proposed that the co-existence of a Compton thick torus and  Compton thin material, extended on a much larger scale, better accounts for the observed phenomenology in absorbed AGN (e.g. Matt 2000).
In addition,  some  Compton thick objects exhibit no optical
activity and a weak scattering X-ray component, indicating that the torus is geometrically
thicker than previously thought (Ueda et al. 2007).
These absorbing media are also responsible for the reflected features (lines and continuum) observed in the X-ray energy domain: a Compton hump above 10 keV and a Fe line at 6.4-6.9 keV (Perola et al. 2002).
Also, the physical/geometrical properties of environmental materials are subjected to the luminosity of the central supermassive black-hole (SMBH) (Bianchi et al. 2007).
On the other hand, it is also important to understand  the  behaviour of the complex multi-component spectrum of an AGN to clarify the physical view of the central engine. To this end,  correct
modelling of the X-ray continuum emission is mandatory. 
Finally, the study of obscured AGN represents a powerful tool to trace the growth of SMBH and to understand how  accretion took place throughout cosmic history. In particular, Compton thick AGN are the best candidates to explain the unresolved fraction of the Cosmic X-ray Background (CXB) in the X-ray energy domain around 30 keV (Gilli et al. 2007).
In this respect, hard X-ray observations, above 10 keV, represent a major improvement 
in selecting obscured AGN that are biased by absorption much less severely than those selected in the 2-10 keV band.
However, because of the limited sensitivity of current X-ray detectors at such high energies, these studies can
only be conducted in the very local Universe.

This kind of study has been performed in a statistical way with a variety of different samples over a broadband energy range: BeppoSAX (Risaliti 2002), \integral\ and \xmm\ (De Rosa et al., 2008), \suzaku\ and Swift-BAT (both type 1 and type 2 AGN, Eguchi et al. 2009, Winter et al. 2009, Fukazawa et al. 2011). 
In the hard X-ray energy range, the IBIS instrument (Ubertini et al. 2003) on board INTEGRAL (Winkler et al.
2003) and the Burst Alert Telescope (BAT, Barthelmy et al.
2005) on board Swift (Gehrels et al. 2004) are currently performing  surveys of the hard X-ray sky, discovering large numbers of sources seen for the first time at these energies (Malizia et al. 2009, Landi et al. 2010, Tueller et al. 2010, Winter et al. 2009, Bassani et al. 2006).
The 4th INTEGRAL/IBIS catalogue (Bird et al. 2010) includes more than 700 hard X-ray sources detected
over a large part of the sky\footnote{70\% of the \integral\ hard X-ray  sky around the Galactic
plane region is covered to better than 1 mCrab sensitivity ($\sim$ 2$\times$10$^{-11}$erg cm$^{-2}$s$^{-1}$),
compared to the 5 mCrab level of 90\% of the extragalactic regions (see details in Bird et al. 2010).} above 20 keV
 and with positional accuracies ranging between 0.2 and 5.5 arcmin, depending on the source significance.


To obtain the best quality broadband high energy data, the \integral/AGN team has undertaken a programme of  follow-up observations of hard X-ray selected AGN with \xmm\ (Molina et al. 2008, Panessa et al. 2011)

In this paper we report on the broadband spectral properties of all type 2 AGN included in the  hard X-ray selected AGN complete sample as  defined by Malizia et al. (2009).
This paper is the continuation of the work we presented for a smaller sample of 7  hard X-ray selected type 2 AGN (De Rosa et al. 2008).
The actual sample comprises 33 AGN optically classified as type 1.8-2.
The new 15 sources detected for the first time above 20 keV with \integral\, have been observed within our \xmm\ follow-up campaign, while we make use of published data to characterise the broadband spectra of 18 previously known objects.

The main goals of this study are (1) to measure
the spectral properties of obscured AGN: the intrinsic continuum shape (both photon index and high energy cut-off) together with the absorption distribution and (2) to
accurately constrain the strength of the reflection components: the Fe K line and the Compton hump. The line properties tell us about the obscuring/reflecting medium while the Compton continuum
is an important yet uncertain parameter (like the high energy cut-off) in population
synthesis models of the CXB (Ueda et al. 2003, Gilli et al. 2007).

The sample is defined in Sect. \ref{sample}, details of the \integral\ and \xmm\ observations are given in 	Sect. \ref{observations}, while the broadband spectral analysis is described in Sect. \ref{spectra}. The results are presented and discussed in Sect. \ref{results}, while  in Sect. \ref{ty1_ty2} we investigate the accretion properties of the sample compared to a sample of type 1 hard X-ray selected AGN. Our results are summarized in Sect. \ref{conclusion}.

\section{The sample}
\label{sample}
From the list of INTEGRAL selected AGN  included in the 3rd IBIS survey (Bird et al. 2007), a complete sample has been extracted by means of the V/V$_{max}$ test, which was first introduced by Schmidt (1968) as a
test of uniformity of the distribution in space for a flux-limited sample of objects (see Malizia et al. 2009 for details).
There are 88 objects detected in the 20--40 keV  band forming this complete sample
of INTEGRAL selected AGN: 46 objects are of type 1 (Seyfert 1--1.5, of which 5 are Narrow Line Seyfert 1s) and 33 of type 2 (Seyfert 1.8-2); only 9 Blazars (BL Lac-QSO) are included in the catalogue.
The list of objects we discuss in this paper is composed of all the type 1.8-2 objects  in the INTEGRAL complete sample of AGN defined in Malizia et al. (2009) except for NGC 1275 and NGC 5506: the first source has been excluded because of the Perseus cluster contamination in the low energy band, while the second source has been re-classified as a Narrow Line Seyfert 1 (see Panessa et al. 2011). To the present sample we have added two sources (IGR J00040+7020 and IGR J20286+2544) 
because they have been observed with moderately long exposure with \xmm\ and are hard X-ray \integral\ selected.

All sources are optically classified, and their redshift distribution has an average value of 0.03 (z$_{max}$=0.252). 
Within our complete sample, 15 are IGR sources, i.e. AGN first detected by INTEGRAL above 20 keV, while 18 are already known type 2 Seyfert (hereafter ''known'').
For the newly discovered objects we started an \xmm\ Large Programme (PI. L. Bassani) with the primary goal to characterize their broadband spectra. The last \xmm\ observation was performed in February 2011.
For the known AGN we used results of broadband studies already published, when available, with Suzaku, \integral\ and \xmm, \integral\ and \chandra\ or BeppoSAX. When broadband data was completely missing from the literature, we analysed unpublished  \xmm\ and \integral\ data.
In conclusion, for this work we performed data analysis and  a spectral study of all IGR sources plus 2 known objects that had no  broadband study already published (NGC4507 and NGC7172) with the \xmm\ data.
The sample, together with additional information (observation dates, exposures in the low and high energy domains,  2--10 and 20--100 keV counts/s and, for the known sources, the instrument used with relative reference), is  presented in Table \ref{journal} where the "new" IGR sources are described in the upper part, and the already "known"  sources in the bottom part.

\section{Observations and data reduction}
\label{observations}

\begin{table*}
\begin{flushleft}
\caption{Journal of the observations}
\begin{tabular}{ccccccc}
\noalign{\hrule}
\noalign{\medskip}

\hline
{Name} & z &  $^\star$(RA,DEC)& Date & Exp Time & Cts(2-10 keV) & Cts(20-100 keV)\\
       &       &  &  XMM obs    & XMM-pn/IBIS & (s$^{-1}$) & (s$^{-1}$) \\
IGR J &  &(J2000)& & (ks) & & \\
\hline
\hline 
00040+7020 & 0.096 & 00 04 01.93 ;70 19 18.6 & 2008-12-29 & 18/2614 & 0.326$\pm$0.004 & 0.23$\pm$0.02 \\
07565-4139 & 0.021 &07 56 19.64 ;-41 37 42.1 &2008-11-18 & 17/1582 & 0.328$\pm$0.005& 0.20$\pm$0.03\\
09523-6231 &  0.252& 09 52 17.00 ;-62 31 00.0 & 2009-01-20&17/1607 &0.321$\pm$0.005 & 0.20$\pm$0.03\\
1009-4250 & 0.033 & 09 52 17.00 ;-62 31 00.0& 2007-06-14 & 21/1020& 1.936$\pm$0.003 & 0.40$\pm$0.04 \\  	
10404-4625 & 0.0239 &10 40 22.31;-46 25 26.5 	 & 2006-11-29 & 11/781 & 1.23$\pm$0.01 &0.48$\pm$0.05 \\		
12026-5349& 0.03 & 12 02 47.68 ;-53 50 08.3 & 2009-12-30 & 25/1805 &0.726$\pm$0.005 &0.60$\pm$0.03 \\ 
13091+1137(4992) & 0.03& 13 09 05.58 ;11 38 02.7 	 & 2006-06-27& 13/495 &0.106$\pm$0.003 &0.51$\pm$0.06 \\
14515-5542 & 0.018 &14 51 33.16;	-55 40 38.8 &2010-02-01 & 18/2617 & 0.555$\pm$0.006 &0.28$\pm$0.03 \\
16024-6107 & 0.011& 16 01 48.23;	-61 08 54.7 	 &2009-09-09 & 19/2022 & 0.358$\pm$0.005 &0.20$\pm$0.03 \\		
16351-5806(137-G34) & 0.0091 & 16 35 10.40 ;-58 06 04.0  &2006-02-13 &17/2351 & 0.039$\pm$0.002& 0.27$\pm$0.03\\
17513-2011 & 0.047 &17 51 13.62 ;-20 12 14.6&2006-09-04 & 8/5980 & 0.567$\pm$0.009 &0.42$\pm$0.02 \\		
20186+4043 & 0.0144 & 20 18 38.50 ;40 41 00.0 &2010-11-01& 12/1202 &0.434$\pm$0.006 & 0.29$\pm$0.03 \\
20286+2544 & 0.01 &20 28 35.06 ;25 44 00.2&2006-04-23 & 11/469& 0.142$\pm$0.004 & 0.53$\pm$0.04\\
23308+7120 & 0.037 & 23 30 37.25 ;71 22 44.8 	 & 2011-01-23 & 33/1995 & 0.162$\pm$0.002 &0.14$\pm$0.03 \\
23524+5842 & 0.164 & 23 52 27.00 ;58 42 00.0  & 2011-02-20 & 18/3498 & 0.232$\pm$0.004& 0.17$\pm$0.02\\	
\hline
known & & & $^{\ddagger}$Date & &  $^\dagger$Instrument & Reference \\
\hline
NGC788 & 0.0136 & 02 01 06.45,;-06 48 57.0 & 1999-01-16 & 39/594 & Chandra -- \integral & De Rosa et al. 2008\\
NGC1068 & 0.038 & 02 42 40.77 ; -00 00 47.8 	 &1996-12-30 & 95/65 & \sax & Matt et al. 1997 \\
NGC1142 & 0.0288 &02 55 12.27 ; -00 11 01.8 	 & 2007-01-23 & 102/81 & Suzaku& Eguchi et al. 2009\\
Mkn3 & 0.0135 & 06 15 36.46 ; 71 02 15.2 & 2005-10-23 & 87/81& Suzaku & Awaki et al. 2008 \\
MCG-05-23-16& 0.0085  & 09 47 40.17; -30 56 55.9 & 2001-12-07 & 98/71 & Suzaku & Reeves et al. 2007 \\
NGC3281 & 0.0115 & 10 31 52.09 ; -34 51 13.4 & 2000-05-20& 71/35 & \sax & Vignali \& Comastri 2002 \\
NGC4388 & 0.0084 & 12 25 46.82; 12 39 43.4	& 2001-06/07& 13/512 & \xmm-- \integral & Beckmann et al. 2004 \\
NGC4507 & 0.0118 & 12 35 36.62 ; -39 54 33.2 & 2001-04-04 & 36/393 & \xmm -- \integral & this work \\
LEDA170194 & 0.036 & 12 39 06.30 ; -16 10 47.2 & 2005-07-25 & 3/200 & Chandra --\integral & De Rosa et al. 2008 \\
NGC4945 & 0.0019 & 13 05 27.28 ; -49 28 04.4 & 2006-01-15 & 99/69& Suzaku & Itoh et al. 2008 \\
CenA & 0.0018 & 13 25 27.62 ; -43 01 08.8 	& 2005-08-19 & 70/60& Suzaku & Markowitz et al. 2007 \\
NGC5252 & 0.023 & 13 38 15.87; 04 32 33.0 & 1998-01-20 &61/29 &\sax &Risaliti 2002 \\
Circinus  & 0.0014 & 14 13 09.91 ; -65 20 20.5 	 & 1998-03-24 &137/63& \sax & Matt et al. 1999  \\
IC4518A	& 0.0163 & 14 57 40.50 ; -43 07 54.0 	 & 2006-08-07 & 11.4/898 & Chandra -- \integral & De Rosa et al. 2008 \\
NGC6300 & 0.0037 & 17 16 59.47 ; -62 49 14.0 & 2001-1999 & 44/77 & \xmm -- \sax/PDS & Matsumoto et al. 2004 \\
ESO103-G35 & 0.0133 & 18 38 20.36 ; -65 25 39.2& 2002-03-15 & 12/44 & Chandra-- \integral & De Rosa et al. 2008 \\
CygA & 0.0561 & 19 59 28.36 ; 40 44 02.1&2000-05-21 & 9/29 & Chandra -- RXTE/PCA & Young et al. 2002 \\
NGC7172 & 0.0087 & 22 02 01.90 ; -31 52 11.6 &2007-04-24 & 28/250 & \xmm -- \integral & this work \\
\hline
\hline
\end{tabular}
\label{journal}
\end{flushleft}
\small{$^\star$ Optical position in hh mm ss.ss; dd mm ss.s. $^\dagger$For Suzaku observation exposures refer to XIXS and PIN data; for \sax\ observation exposures refer to MECS and PDS data; for Chandra observations exposures refer to ACIS data; for \xmm\ observation refer to pn data; for RXTE to PCA data. $^{\ddagger}$ Date of observations performed with \integral\, and  \xmm\ or \chandra\ refer to \xmm/\chandra\ observations.}
\end{table*}

\subsection{INTEGRAL/IBIS}

Though the sources reported here have been hard X-ray detected in  the 3rd \integral\ survey (Bird et al. 2007), to better characterize  the spectrum above 20 keV we use the data available in the 4th \integral\ catalogue (Bird et al. 2010), making use of the longer on-source exposures.
We also stress here that we used \integral/IBIS data  processed  by the survey team following the procedures as described below and reported in  Bird et al. (2010).
For each source, they combined \integral/IBIS data  from  several pointings performed between
revolution 12 and 530  to provide sufficient statistics for weak mCrab flux sources.
First  they generated IBIS/\emph{ISGRI} images for each available pointing  in 13 energy
bands with the ISDC offline scientific analysis software OSA
version 7.1 and they then extracted count rates at each source position
from individual images to provide light curves
in the various energy bands sampled; since the light curves did not show any sign of variability or flaring activity, average fluxes were
then extracted in each band and combined to produce the source spectrum. 
In Table \ref{journal}, we report the details of the IBIS/ISGRI observations;
(optical) best fit positions, exposures and count rates in the range 20--100 keV band. \\

\subsection{XMM-Newton}

To avoid  cross-calibration problems and for the sake of simplicity, we discuss in this paper only the data of the EPIC/pn camera (Struder et al. 2001), however  the MOS1 and MOS2
spectra have been checked for consistency. Data were reduced using SAS v10.0 employing \texttt{eproc} with standard settings and the most updated calibration files available at the time of the data reduction (April 2011). Source spectra were extracted from circular regions surrounding the source position while background spectra were extracted in the same CCD chip from circular regions free from contaminating serendipitous sources.
Each spectrum was rebinned in order to have at least 30 cts per bin  allowing us to apply $\chi^2$ statistics. Spectral fits were performed in the 0.3--10 keV energy band.
The information on the \xmm\, observations (date, exposure and 2--10 keV count rate) is shown in Table \ref{journal}.

\section{Spectral analysis}
\label{spectra}

We fitted simultaneously the soft and hard X-ray spectra available with \integral/IBIS and  \xmm\ in the 0.3-100 keV energy range. 
To take into account possible flux variation of the source between the non simultaneous \xmm\ and \integral\ data, we left  a cross-calibration constant, between IBIS and pn, free to vary during the fit procedure.
Possible miscalibration  between \integral\ and \xmm\ could mimic or hide the presence of a Compton reflection component above 10 keV; we will discuss this point in Sect. \ref{reflection}. 

For  the sake of clarity and to ease the comparison with the published results on the known objects, we considered a baseline model (BLM thereafter) which is commonly used to fit the broadband X-ray spectra (0.3--100 ekV) of absorbed AGN (Comastri et al. 2010, Ueda et al. 2007, Risaliti 2002, De Rosa et al. 2008). The BLM is composed of:
1) a primary continuum extending up to the hard X-ray energy domain  modelled by a cutoff power--law having photon index $\Gamma$ and high energy roll-over E$_{c}$;
2) a soft X-ray scattered component modelled by a power--law  having photon index  $\Gamma_{s}$ assumed to be equal to  $\Gamma$ and a scattering fraction f$_{sc}$;
3) a Compton reflection continuum above 10 keV with the inclination angle $\theta$ of the reflector fixed to 45 deg, and solar abundance, plus the Fe K$\alpha$ emission line that is reproduced with a  gaussian profile.
We assume the Fe line to be intrinsically narrow  ($\sigma<$0.01 keV, FWHM$<$1000 km s$^{-1}$) and no additional Fe K$\beta$ emission component is included in this model; this is due mainly to the fact that we are characterizing the continuum properties.    
However, in some cases we need to relax the assumption about narrowness to well reproduce the Fe line; these cases will be discussed in Sect. \ref{feline}.
All the  spectral components are absorbed by the Galactic column density along the line of sight  N$_{H}^{Gal}$\footnote{N$_{H}^{Gal}$ values have calculated through the web interface http://heasarc.gsfc.nasa.gov/cgi-bin/Tools/w3nh/w3nh.pl, using the Dickey \& Lockman (1990) HI in the Galaxy.}, while the primary power--law is also absorbed by a second cold medium with column density N$_{H}$.
We can write the BLM of the photon spectrum F(E) as follows: 
 
$e^{-\sigma_{th}N_{H}^{Gal}} (e^{-\sigma_{th}N_{H}}K E^{-\Gamma}e^{-E/E_{c}}+f_{sc} K E^{-\Gamma_{s}}+R(E)+G(E)+S(E)) $

where G(E) represents the Fe-K Gaussian emission line,  R(E) is the Compton reflection component (with the shape fixed to that of the primary emission), and S(E) represents any additional soft components needed to reproduce the soft X-ray spectra (see Sect. \ref{soft X-ray}).
The total baseline model (excluding Galactic absorption) is expressed as \texttt{zphabs*cutoffpl + const*cutoffpl + zgauss + pexrav + \textit{soft comp}} in the XSPEC terminology.
The broadband spectra of IGR sources are shown in Fig.\ref{spec1}, \ref{spec2} and \ref{spec3} while the best fit parameters for the whole sample are reported in Table \ref{mastertab}.

In this paper we quote statistical errors at the 90\% confidence
level for one interesting parameter unless otherwise
specified. In order to calculate the luminosities we use a standard $\Lambda$ cold dark matter cosmology with the following
cosmological parameters: H$_{0}$=70 km s$^{-1}$ Mpc$^{-1}$,
$\Lambda_{0}$=0.73, $\Omega_{M}$=0.27 (Bennett et al. 2003).

\begin{table*}
\begin{flushleft}
\caption{Broadband baseline model (BLM) composed by an absorbed power--law with exponential cutoff+reflection continuum+ Fe line + soft scattered component with $\Gamma_{soft}$=$\Gamma$}
\begin{tabular}{ccccccccccc}
\noalign{\hrule}
\noalign{\medskip}
{Name} & $^1\Gamma$ & $^2$N$_{\rm H}$& $^3$R & $^4$E$_{\rm c}$& $^5$f$ _{\rm sc} $ & $^6$EW & $^7$F$ _{\rm 2-10 keV} $ & $^8$F$ _{\rm 20-100 keV} $ & $^9$log L$^{corr}_{(2-10 keV)}$ & $^9$log L$_{(20-100 keV)}$\\	
IGR J &   & & & &   & &  &  & &\\	
\hline 
00040+7020 & $ 1.44^{+0.06}_{-0.02} $ &  $ 33^{+2}_{-2} $ & $ <1.8$ & $ >120 $  & $ 0.9^{+0.5}_{-0.5} $ & $ 50^{+48}_{-47} $ & 0.35 & 1.6 & 44 & 44.6 \\	
07565-4139 &  $ 1.79^{+0.08}_{-0.03} $ & $ 7.2^{+0.4}_{-0.9} $ & $ 3.5^{+0.6}_{-1.3} $&  $ >100 $ & $ 5^{+4}_{-2} $ & $ 163^{+113}_{-113} $ &0.32 &  1.3 & 42.5 & 43.1\\	
09523-6231 &   $ 1.75^{+0.09}_{-0.09} $& $ 63^{+4}_{-4} $&$<$2 &$>$40 & $1.8^{+0.3}_{-0.3} $& $ 77^{+35}_{-35} $& 0.37  & 1.3 & 45 &45.5 \\	
1009-4250& $ 1.66^{+0.04}_{-0.04} $ & $ 257^{+15}_{-14} $&$ 1.4^{+0.3}_{-0.3} $ & $ >150 $ &$ 1.0^{+0.1}_{-0.1} $ & $ 70^{+48}_{-25} $& 0.31&2.7 & 43.3 &43.9\\	
10404-4625 &$ 2.17^{+0.03}_{-0.03} $ & $ 41^{+1}_{-1} $& $ 4.3^{+1.4}_{-0.5} $ & $ >250 $ & $<0.2$ & $ 189^{+30}_{-65} $&1.2 &2.7& 43.3& 43.6\\	
12026-5349&$ 1.16^{+0.05}_{-0.03} $ & $ 29^{+2}_{-2} $ & $ 1.1^{+0.5}_{-0.5} $  & $ 76^{+14}_{-11} $  & $ 4.1^{+0.1}_{-0.1} $ & $ 100^{+20}_{-50} $& 0.85 &3.7&43.3&43.8 \\	
13091+1137 & $ 1.15^{+0.09}_{-0.09} $ & $ 432^{+36}_{-36} $ & $ <0.4 $& $107^{+193}_{-20}$ & $ 0.5^{+0.1}_{-0.1} $& $ 379^{+81}_{-81} $& 0.21& 3.8 & 43.1& 43.8\\	
14515-5542 & $ 1.64^{+0.11}_{-0.08} $&  $ 3.3^{+1.7}_{-0.8} $ & $ 1.4^{+1.9}_{-1.0} $ & $ >80$  &$ 30^{+10}_{-20} $ & $ 182^{+46}_{-46} $ & 0.53 & 1.9 & 42.6 & 43.2\\	
16024-6107 &  $ 2.26^{+0.08}_{-0.10} $ & $ 2.5^{+0.3}_{-0.1} $ &$ 3.3^{+1.8}_{-0.6}$ &  $>70$ & $<3.2$ &$ 121^{+80}_{-80} $ & 0.30 & 1.3 & 41.9 & 42.6\\	
16351-5806 & $ 1.2^{+0.1}_{-0.1} $ & $ 2000^{+10}_{-10} $ & $0.8^{+0.2}_{-0.2}$ &  $ >150 $ & $9.0^{+0.6}_{-1.0}$ & $ 1400^{+300}_{-300} $ &0.05 &1.5&42.1&42.5\\	
17513-2011 & $ 1.7^{+0.2}_{-0.2} $& $ 6.6^{+0.1}_{-0.1} $ & $ 3.5^{+3.8}_{-1.7} $  & $ >150$ & $<12$ & $ <200 $& 0.58&3.0&43.5& 44.2\\	
20186+4043 & $ 1.6^{+0.3}_{-0.2} $ & $ 58^{+5}_{-5} $  & $2.0^{+1.3}_{-0.9}$ & $50^{+112}_{-38}$ & $ 1.9^{+0.9}_{-0.4} $ & $ 80^{+80}_{-37} $& 0.53 & 2.0 & 42.6&42.9\\	 
20286+2544 & $ 1.4^{+0.1}_{-0.1} $& $ 574^{+96}_{-66} $ & $<1$ & $ >180$ & $ 0.5^{+0.2}_{-0.2} $& $ 112^{+56}_{-56} $ & 0.28 & 5.5 & 42.8&43.3\\	
23308+7120 &  $ 2.0^{+0.2}_{-0.4} $ & $ 91^{+17}_{-12} $ & $ <2.7 $ & $ 63^{+340}_{-40} $ & $<2$ &$100^{+50}_{-50} $ & 0.14 & 0.9 & 42.9&43.4\\	
23524+5842 &  $ 1.5^{+0.1}_{-0.1} $& $ 29^{+4}_{-3} $& $<$3 & $>$120&$ 5^{+2.0}_{-2.0} $ &$ 100^{+60}_{-60} $& 0.28 & 1.1 & 44.4 & 44.9\\	
\hline
known &   & & & &   & &  &  & &\\	
\hline
NGC788 & $ 1.25^{+0.05}_{-0.17} $ & $ 300^{+60}_{-30} $ & $ 0.9^{+1.3}_{-0.7} $ & $ 62^{+38}_{-24} $ & -- &$ 539^{+231}_{-109} $ & 0.61 & 3.8 & 42.8 & 43.1\\
NGC1068 &  $ 2.13^{+0.17}_{-0.17} $ & $ 1e4 $ & - & -- & $<$2 &$ 1000^{+140}_{-140} $ & 0.52 & 3.9 & 44 &41.9 \\
NGC1142 &   $ 1.78^{+0.02}_{-0.02} $ & $ 631^{+21}_{-18} $ &  $ 1.49^{+0.15}_{-0.23} $& $>300$  &$ 0.26^{+0.05}_{-0.05} $ & $218^{+20}_{-20}$ &0.42 & 5.3 & 43.7&44.0\\
Mkn3 &  $ 1.80^{+0.06}_{-0.06} $ & $ 1000^{+60}_{-60} $ & $ 1.36^{+0.06}_{-0.11} $ & $>200$ &$0.9^{+0.2}_{-0.2} $ & $855^{+32}_{-32}$&  0.64&10&43.2&43.6\\
MCG-05-23-16 &  $ 1.95^{+0.03}_{-0.03} $ & $ 16.5^{+0.3}_{-0.3} $ & $ 1.1^{+0.2}_{-0.2} $& $>170$ & $ 0.50^{+0.01}_{-0.01}$& $ 70^{+6}_{-6} $ &8.76 &14.5 & 43.2&43.5 \\
NGC3281 & $ 1.95^{+0.18}_{-0.18} $ &  $ 1500^{+200}_{-200} $ & $ 0.3^{+0.1}_{-0.1} $ & $>60$  & -- &$ 1180^{+400}_{-361} $ &0.29 &7.2 & 43.2&43.3\\
NGC4388 & $ 1.65^{+0.04}_{-0.04} $& $ 273^{+7}_{-7} $& $ 0.3^{+0.2}_{-0.1} $ & $>180$ & -- & $ 220^{+10}_{-10} $& 2.3 & 20 & 43&43.6\\
NGC4507 & $ 1.80^{+0.14}_{-0.17} $ & $ 440^{+54}_{-57} $ & $ 1.5^{}_{} $& $ 126^{+153}_{-48} $ & $ 1.5^{}_{} $ &$ 117^{+27}_{-27} $ &1.3 &17.4&43.2&43.8 \\
LEDA170194  &  $1.7^\star$ & $ 29^{+13}_{-3} $ & $ <$4.8 & $>$ 210 &-- &$<3000$ & 2.0 & 4.9 & 43.9&44.2\\
NGC4945 & $ 1.6^{+0.1}_{-0.2} $ & $ 5300^{+400}_{-900} $ & -- & $>80$ & -- &$ 1600^{+300}_{-300} $ &0.54 &28.9 & 43 &42.6 \\
CenA & $ 1.82^{+0.02}_{-0.01} $ &  $ 147^{+3}_{-2} $& $ <0.2$ &$>400$  & $<$1.6 &$ 83^{+3}_{-3} $ &21.2 & 64 & 41.9 &43.0\\
NGC5252 & $ 1.83^{+0.13}_{-0.12} $ &  $ 68^{+16}_{-7} $& $ 1.6^{+1.2}_{-1.2} $ &$>50$  & -- &$ 278^{+215}_{-171} $ &0.26 & 0.66 & 42.7&42.9 \\
Circinus  &  $ 1.56^{+0.16}_{-0.51} $ &  $ 4300^{+400}_{-700} $& $ 0.2^{+0.1}_{-0.1} $ & $ 56^{+8}_{-23} $ & $ 0.8^{+0.1}_{-0.1} $& $ 2250^{+100}_{-100}$& 1.4 & 23 & 42.2 & 42.3 \\
IC4518A	&  $ 1.5^{+0.2}_{-0.3} $ & $ 140^{+30}_{-10} $ & $ 2.6^{+3.5}_{-2.5} $ & $ 70^{+60}_{-30} $ & -- &$ 360^{+106}_{-190} $ & 0.29 &2.9 & 42.6 &43.3\\
NGC6300 &  $ 1.94^{+0.09}_{-0.09} $ & $ 240^{+10}_{-20} $ & $ 1.1^{+1.2}_{-0.6} $ & $ > $250 & 0.2 &$ 140^{+50}_{-50} $ &0.86 &9.9& 41.8 &42.1\\
ESO103-G35 &  $ 1.9^{+0.2}_{-0.2} $& $ 189^{+6}_{-11} $& $ 1.7^{+0.6}_{-0.5} $  & $ 50^{+250}_{-25} $  &  -- & $ 325^{+100}_{-100} $ & 2.4 &7.9 & 43.4&43.5 \\			
CygA &  $ 1.52^{+0.12}_{-0.12} $ &  $ 200^{+10}_{-20} $& $ -- $ & --  &$ 1.1^{+0.1}_{-0.1} $   &$ 182^{+43}_{-43} $ & 1.2 & 4.0 & 44.6&44.5\\
NGC7172 &  $ 1.58^{+0.01}_{-0.01} $ & $ 73.8^{+0.7}_{-0.7} $ & $ 0.3^{+0.1}_{-0.1} $ & $ 66^{+12}_{-10} $ &$ 0.20^{+0.01}_{-0.03} $ &$ 76^{+10}_{-10} $& 4.3 &8.2 & 43 &43.0\\
\hline
\end{tabular}
\label{mastertab}
\end{flushleft}

\small{$^\star$ frozen in the fit;  $^1$ photon index of the primary absorbed, power--law; ; $^2$absorber column density in 10$^{21}$ cm$^{-2}$; $^3$ Relative reflection; $^4$ high energy cutoff in keV; $^5$ scattered fraction; $^6$ Fe K$\alpha$ equivalent width measured in eV with respect to the observed total continuum; $^7$ Observed flux in the 2--10 keV energy range in 10$^{-11}$ erg cm$^{-2}$s$^{-1}$; $^8$Observed flux in 20-100 keV energy range in 10$^{-11}$ erg cm$^{-2}$s$^{-1}$;  $^9$ Log of the luminosity in 2--10 keV energy range corrected for absorption in erg s$^{-1}$;}
\end{table*}

\section{Results and discussion}
\label{results}

\subsection{The primary continuum}
\label{continuum}

\begin{figure}
\centering
\includegraphics[width=0.4\linewidth]{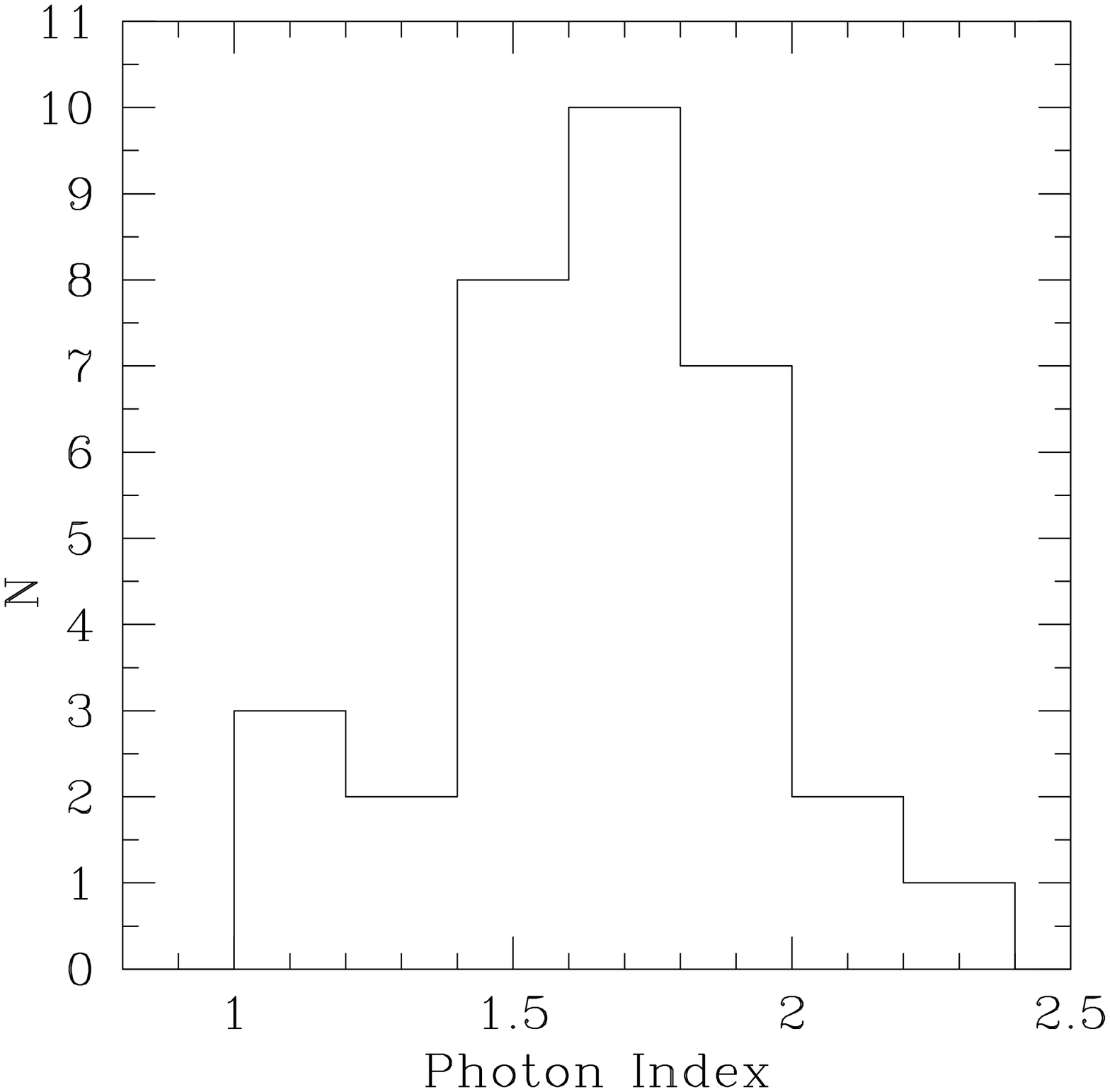}
\includegraphics[width=0.4\linewidth]{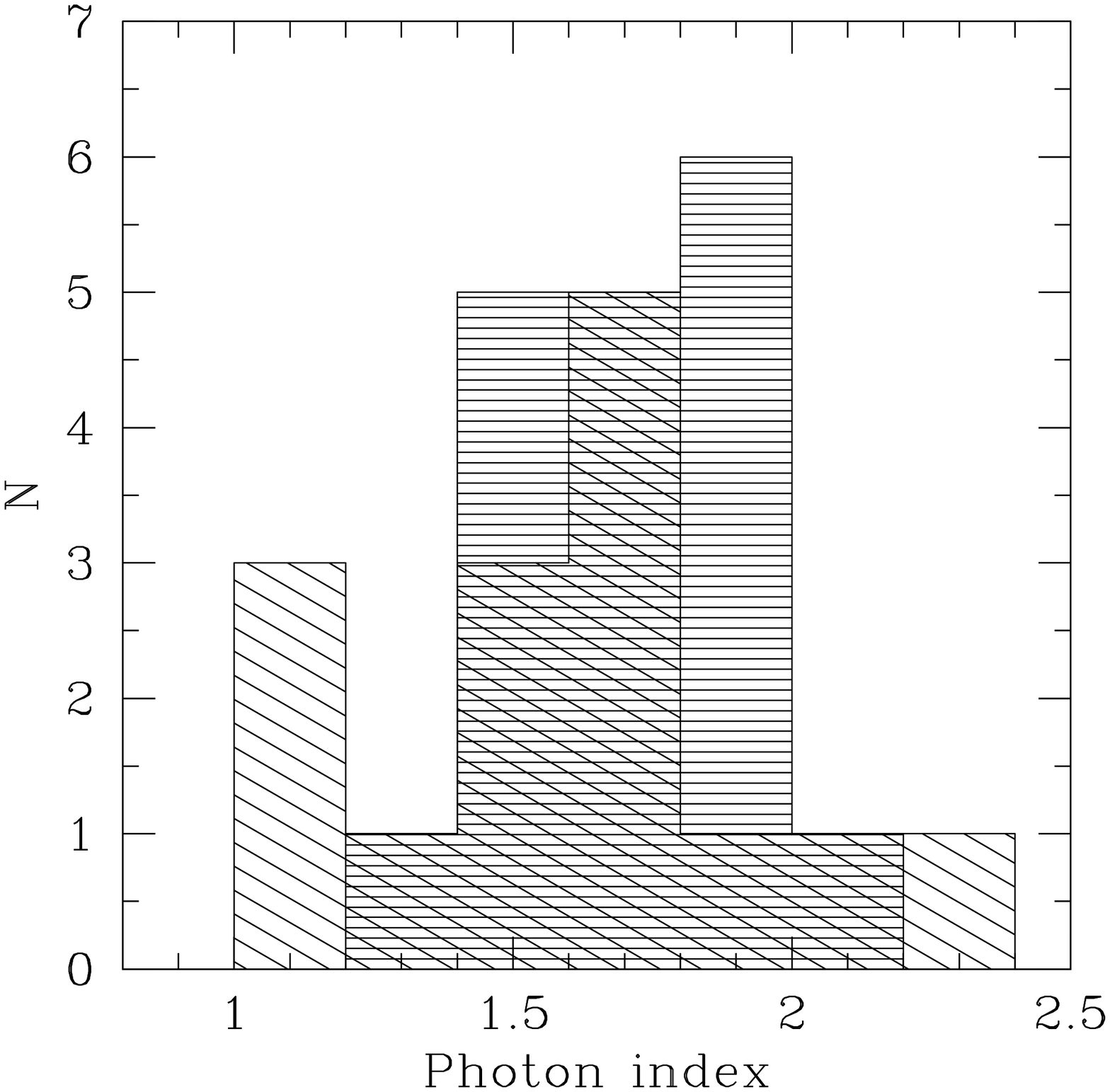}
\caption{\textit{Left panel}: Primary photon index $\Gamma$ distributions for the whole sample. \textit{Right  panel}: Photon index distributions for two separated population: IGR (diagonal dashes) and known literature sources (horizontal dashes).}
\label{pho_distr}
\end{figure}

\begin{figure}
\centering
\includegraphics[width=0.4\linewidth]{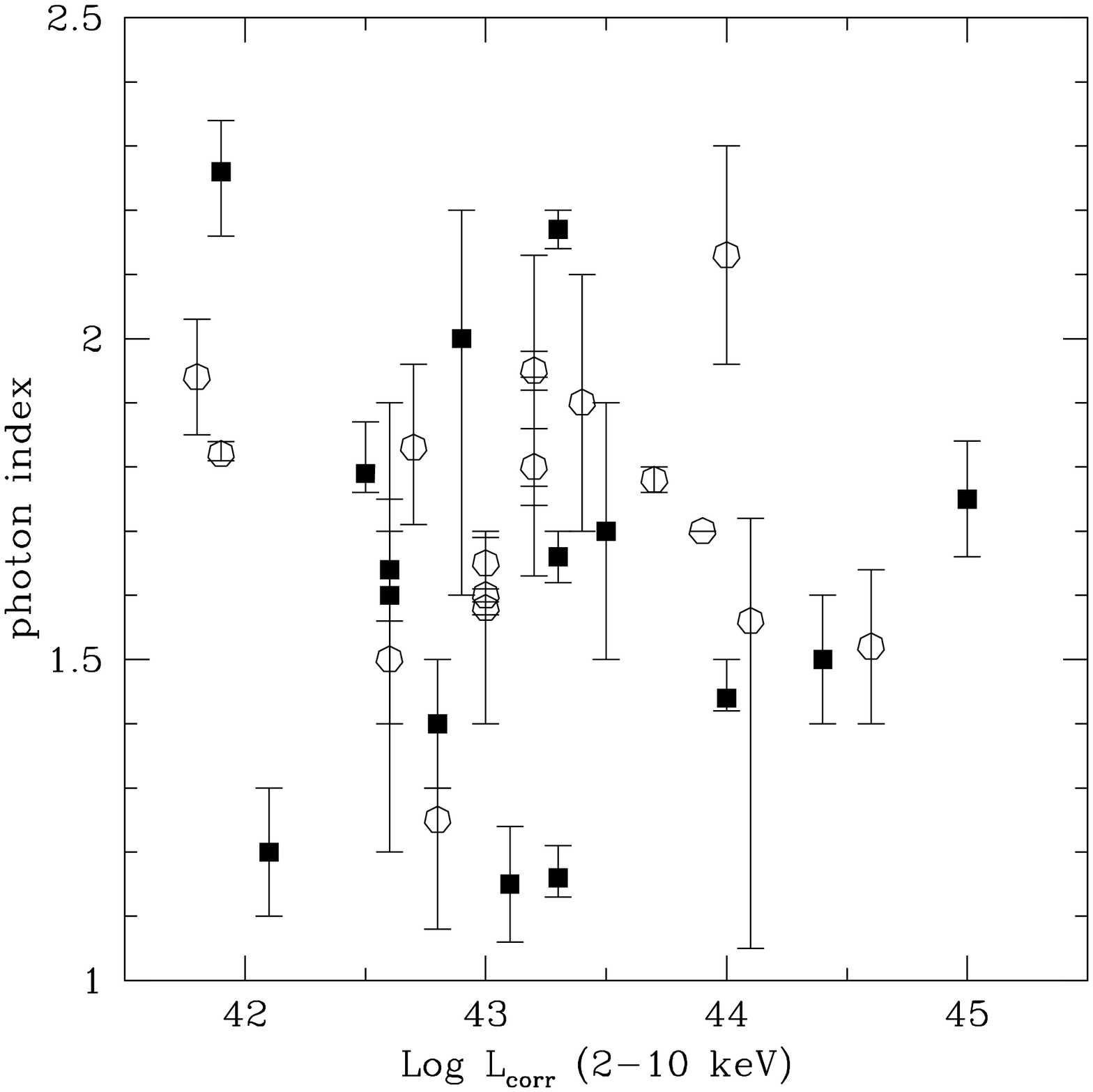}
\includegraphics[width=0.4\linewidth]{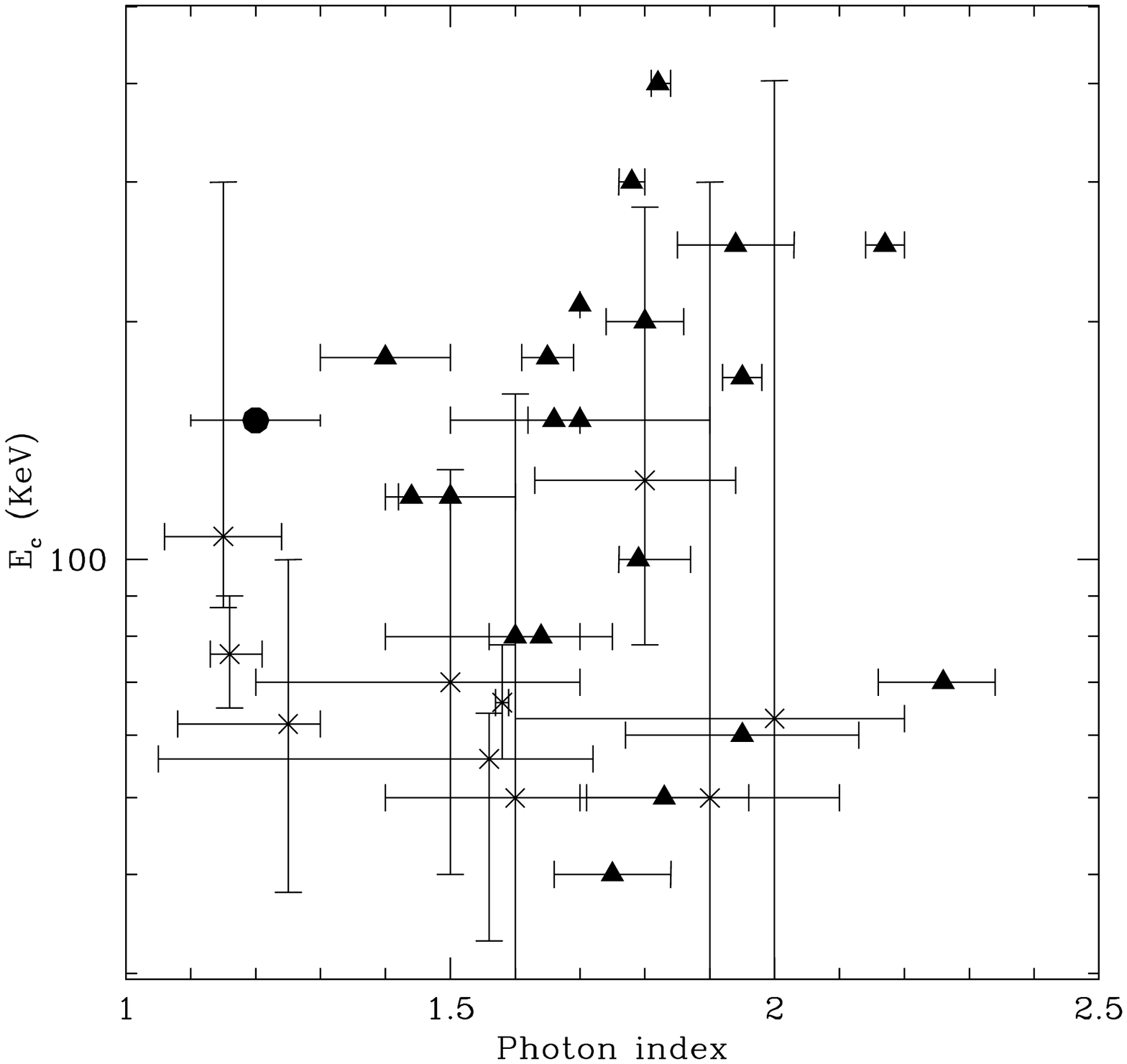}
\caption{\textit{Left panel}: photon index for the whole sample as a function of the 2--10 keV luminosity corrected for absorption. Filled boxes are IGR sources and open circles are known sources. \textit{Right panel}: 
 High-energy cutoff vs photon index for the whole sample. Filled triangles are lower limits to E$_c$ while the filled dot is the upper limit of IGR J16351-5806.}
\label{cutoff-lum}
\end{figure}

The BLM  reproduces quite well all the spectra analysed since the $\chi^{2}$/dof ranges between 0.77-1.12.
In Fig. \ref{pho_distr} we show the distribution of the photon index of the primary absorbed power--law for the whole sample (on the left) as well as  for the IGR and known objects separately (on the right) to search for possible differences in the radiative emission mechanisms  between the two sample of sources.
The mean value of the photon index and its standard deviation for the whole sample is $\Gamma$=1.68, $\sigma$=0.30. This value is in very good agreement with those found in previous analyses performed in a limited energy range (2--10 keV) in both type 1 and type 2 AGN (e.g, Singh et al. 2011, Bianchi et al. 2009).  A detailed comparison between type 1 and type 2 Seyfert will be presented in Sect. \ref{ty1_ty2}.
We note that the average value of the IGR population is only marginally flatter than the known one, being $\Gamma_{IGR}$=1.63, $\sigma_{IGR}$=0.34 and  $\Gamma_{known}$=1.74, $\sigma_{known}$=0.21. A Kolmogorov-Smirnov test applied on these two populations indicates that they come from the same parent distribution with a probability of P=0.385.
To explore a possible correlation between the primary continuum and the brightness, we plot in Fig. \ref{cutoff-lum} (left panel) the photon index against the 2--10 keV luminosity corrected for absorption: we do not find any evidence of correlation, a Pearson test for linear fit yields r=-0.02.

We measured the high energy cutoff in 10 out of 33 sources (i.e. 30 per cent of the sample), namely: NGC4507, IGR J12026-5349, IGR J13091+1137, IGR J20186+4043, IGR J23308+7120, IC4518A, NGC7172, Circinus, NGC788 and ESO103-G35, while only an upper limit at 300 keV has been found in IGR J16351-5806. In the remaining sources we obtained lower limits to E$_{c}$. The average value of the ten measured high energy cutoffs is E$_c$=73 keV with a standard deviation $\sigma$=25 keV.
In Fig. \ref{cutoff-lum} (right panel) we plot the high energy cutoffs as a function of the photon index $\Gamma$. All the measured values are below 150 keV, while lower limits (filled triangles) are well grouped below 300 keV. These values are consistent with those observed in the Seyfert 1 by BeppoSAX (Perola et al. 2002, De Rosa et al. 2007). We do not find any evidence of correlation between $\Gamma$ and E$_c$,  the correlation coefficient being r=0.1.
Our analysis strongly suggests that the high energy rollover is an ubiquitous property of Seyfert galaxies, and that its value  ranges in the same intervals for type 2 and type 1 Seyferts (Molina et al. 2008). 
It is worth noting that we were able to measure E$_{c}$ in the objects with the highest S/N ratio, suggesting that future deep hard X-ray observations will be the right way to investigate the presence of this component in the spectra of AGN.

\begin{figure}
\centering
\includegraphics[width=0.4\linewidth]{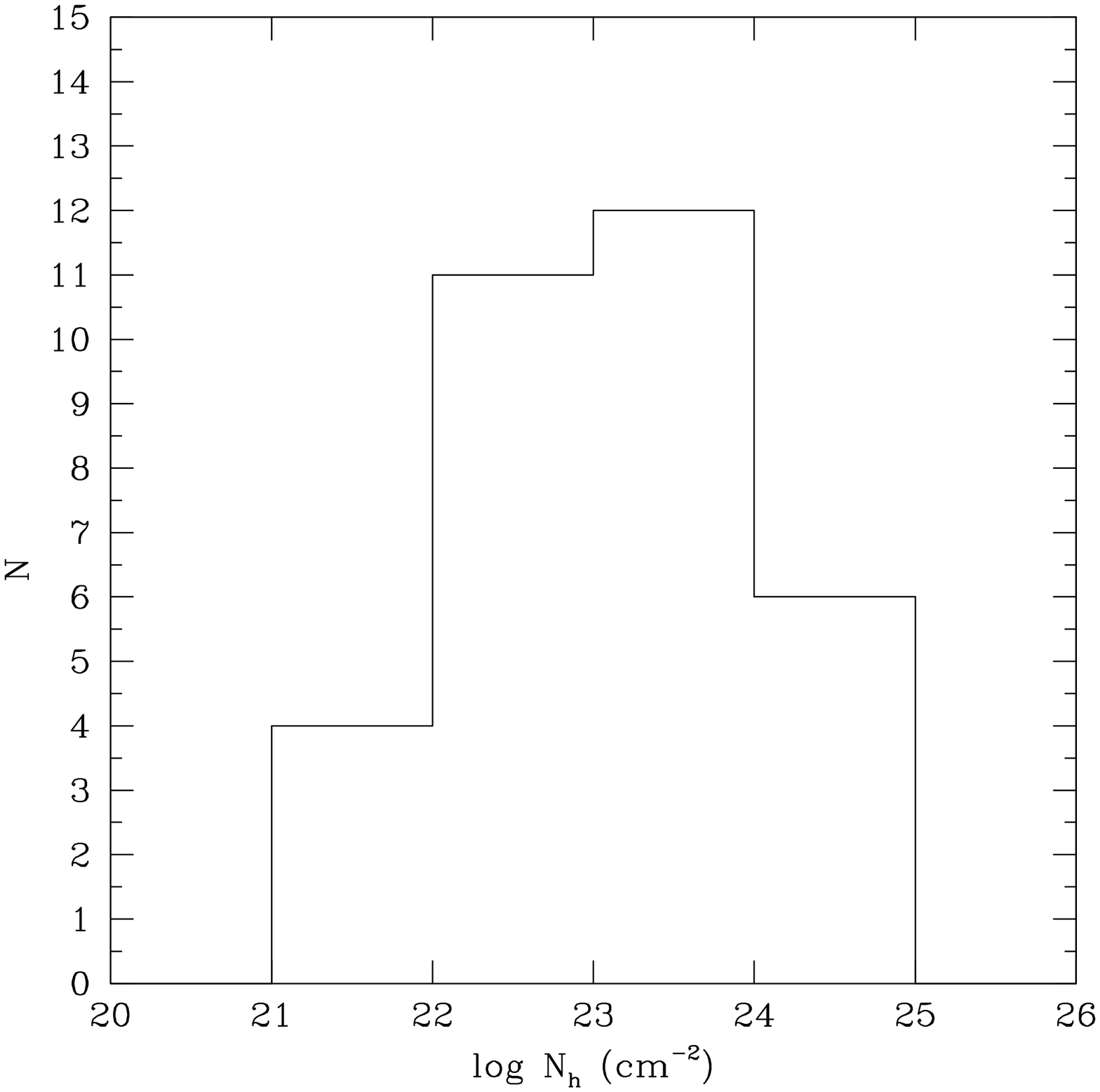}
\includegraphics[width=0.4\linewidth]{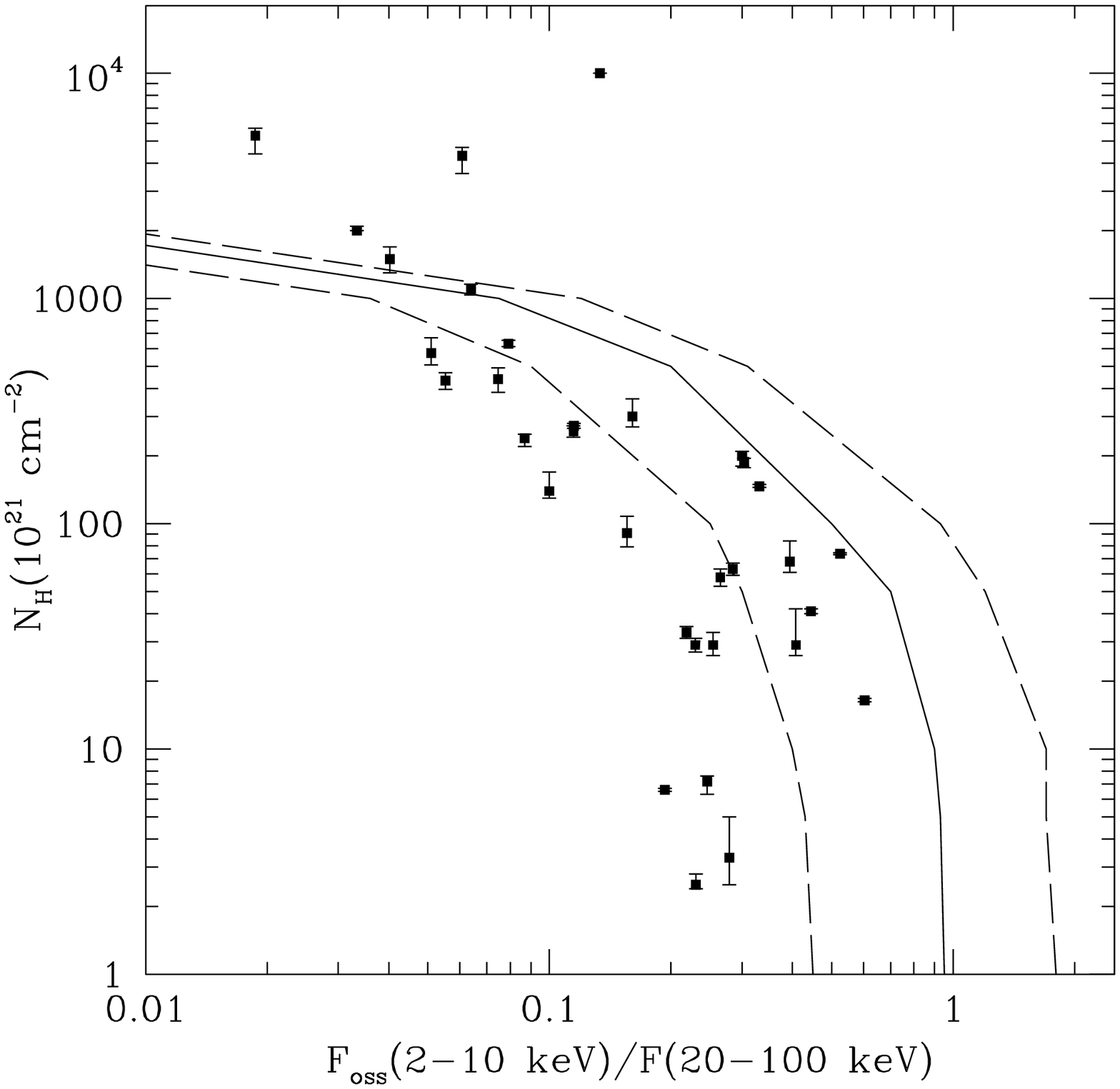}
\caption{\textit{Left panel}: Distribution of the absorbing column density of the whole sample. 
\textit{Right panel}: The column density distribution as a function of the softness ratio F$_{oss}$(2--10 keV)/F(20--100 keV). The lines represent a region in the diagnostic defined when accounting for the typical  values of the photon index and high energy cutoff, with errors, as found in our analysis.} A clear trend is evident with Compton thick sources showing a lower softness ratio.
\label{Nh_distr}
\end{figure}


\subsection{The absorption properties}
\label{absorption}

The absorbing column density distribution of the primary emission is shown in the left panel of Fig.\ref{Nh_distr}. Its average value is 23.15 with a  standard deviation $\sigma$=0.89. Excluding the six Compton thick sources (IGR J16351-5806, NGC 3281, NGC 4945, Circinus, NGC 6300, NGC 1068) we obtain an average value (standard deviation) of 22.85 (0.670). 
The absorbing column density distribution of the whole sample is in full agreement with that reported by Malizia et al. (2009) for the same sample.
When our N$_{H}$ distribution (left panel in Fig. \ref{Nh_distr}) is  compared with the absorption distribution obtained with 2--10 keV data only, e.g. in the \xmm\ and \chandra\ deep surveys (Giacconi et al. 2002, Worsley et al. 2004, Akylas et al. 2006), it becomes evident that the hard X-ray  data strongly favours the detection of more heavily obscured sources than 2--10 keV data only.
Nonetheless, in the hard X-ray domain above 10 keV a significant fraction of the source flux might be lost if the source is heavily Compton thick (log(N$_{H})>$25). In fact, a number of objects of this type is not present in our sample as well as in the \textit{Swift}/BAT survey above 10 keV (Burlon et al. 2011).

In Fig. \ref{Nh_distr}, right panel, we plot the absorbing column density as a function of the ratio of the observed 2--10 keV flux versus the 20--100 keV flux. This diagnostic plot was proposed by Malizia et al. (2007) in an attempt to isolate peculiar objects among  a sample of 34 \integral\ AGN observed with Swift-XRT.
The plot shows a clear anticorrelation between N$_{H}$ and the softness ratio: this ratio becomes lower as the column density increases.
The lines define a region in the plot obtained by assuming an absorbed power-law with high energy cutoff, where we have used  the average values of $\Gamma$ and E$_c$, with errors, as found in our sample (see Sect. \ref{continuum}).
In this diagnostic plot, Compton thick AGN are well isolated, all six Compton thick objects of our sample are located above the line.

\subsection{The reflection continuum}
\label{reflection}

\begin{figure}
\centering
\includegraphics[width=0.4\linewidth]{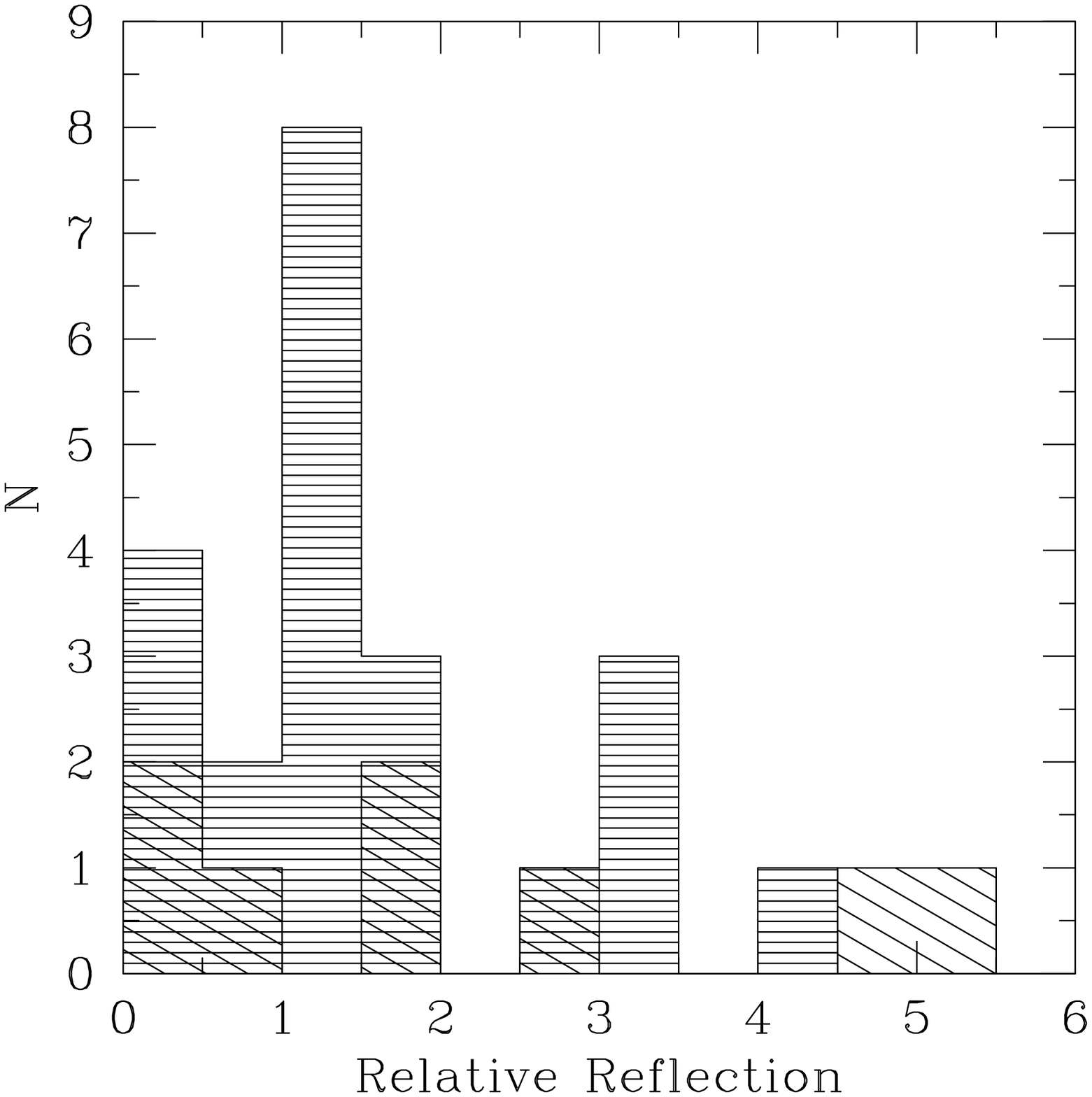}
\includegraphics[width=0.4\linewidth]{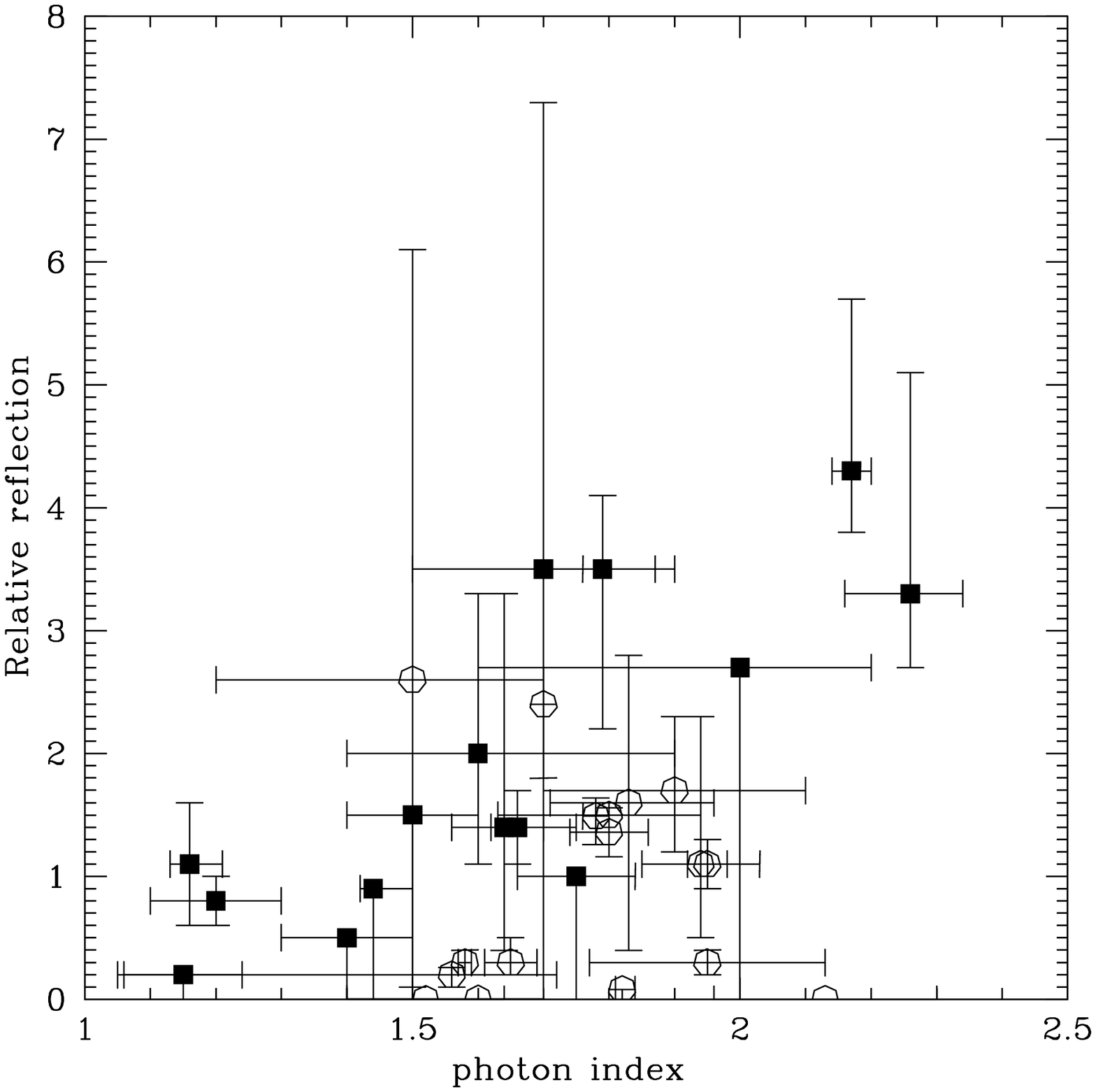}
\caption{\textit{Left panel}: Distribution of the reflection fraction for all sources in the sample. Upper limits are shown whit obliques lines while detections are horizontal lines. \textit{Right panel}: Relative reflection as a function of the photon index, filled boxes represent IGR sources, open circles are known AGN already known.}
\label{refl_distr}
\end{figure}

Our BLM includes the Compton reflection components, namely, the Compton hump and the Fe K$\alpha$ line.
We have  measured the reflection fraction (R) in 22 sources (average value R=1.63 with $\sigma$=1.15), while for 8 objects we only found upper limits.
The distribution of the values of R derived for the whole sample is shown in Fig. \ref{refl_distr} where upper limits are shown as obliques bars while the measured values are filled with horizontal bars.

In the right panel of Fig. \ref{refl_distr} we show the relative reflection R as a function of the photon index for the whole sample.
The data show a poor correlation coefficient (r=0.4), indicating that the correlation claimed by Zdziarski
et al (1999) between R and $\Gamma$ is not found in our data. It is worth noting that the fit of our BLM produces a degeneracy between the reflection fraction R and the photon index $\Gamma$: a  higher $\Gamma$  produces a higher value of R.
 Another point to stress is that the soft and high energy data (below and above 10 keV) are non-simultaneous.
 \integral/IBIS data are the result of the average flux state of the source over a time interval of about 5 years (the time interval covered by the fourth \integral\ catalogue Bird et al. 2010). \xmm\ observations last at most 20 ks and are therefore a snapshot not necessarily simultaneous with  \integral/IBIS pointings (see Table \ref{journal}). To take into account possible flux variation between \xmm\ and \integral\ data we added in the model a normalization constant C. A value of C different from one would indicate a flux variation between the soft and hard-X ray energy domain.
 The derived values of C  are well clustered around 1, with an average value of 1.45 and a standard deviation of 0.76. For two sources only, namely  IGR J16024-6107 and IGR J23308+7120,  C is significantly larger than 1  (3.3$^{+0.9}_{-0.7}$ and 3.1$^{+1.1}_{-1.2}$, respectively), suggesting  flux variability between the soft and hard X-ray observations. To support this conclusion we note that in the \integral/IBIS catalogue (Bird et al. 2010) both have a bursticity index flagged with yes\footnote{The bursticity of a source is defined as
the ratio of the maximum significance on any timescale, compared to the significance defined for
the whole dataset. A flag of yes (Y) indicates a bursticity greater than 1. See Bird et al. 2010 for more details}, and both show flux variation in the 20--40 keV band larger than a factor of three, in their ligh-curves. In addition, in the case of IGR J16024-6107 the high value of R would suggest that this object is peculiar in several aspects.
By excluding these sources from the sample, the average value of C (standard deviation) becomes lower, 1.19 (0.31).
As a result, we are confident that the value of R and E$_{c}$ we found are realistic and not due to variability.

For the purpose of this work we do not examine the source variability in detail  as we are mainly interested in explaining the average properties of the sample. However, to retain the completeness of our sample, we chose to include also the two  variable sources (IGR J16024-6107 and IGR J23308+7120) in our analysis.

\begin{figure}
\centering
\includegraphics[width=0.4\linewidth]{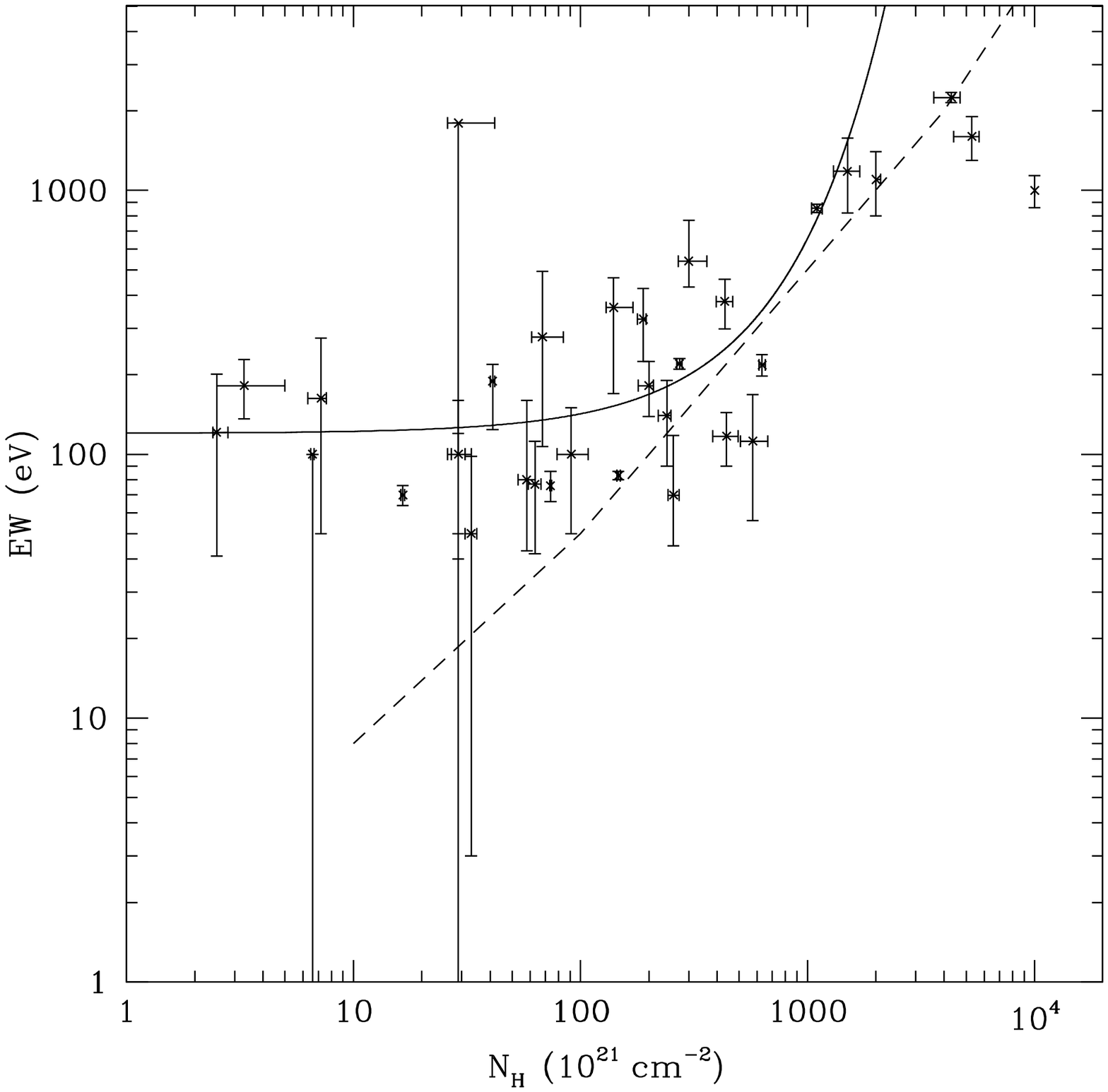}
\includegraphics[width=0.4\linewidth]{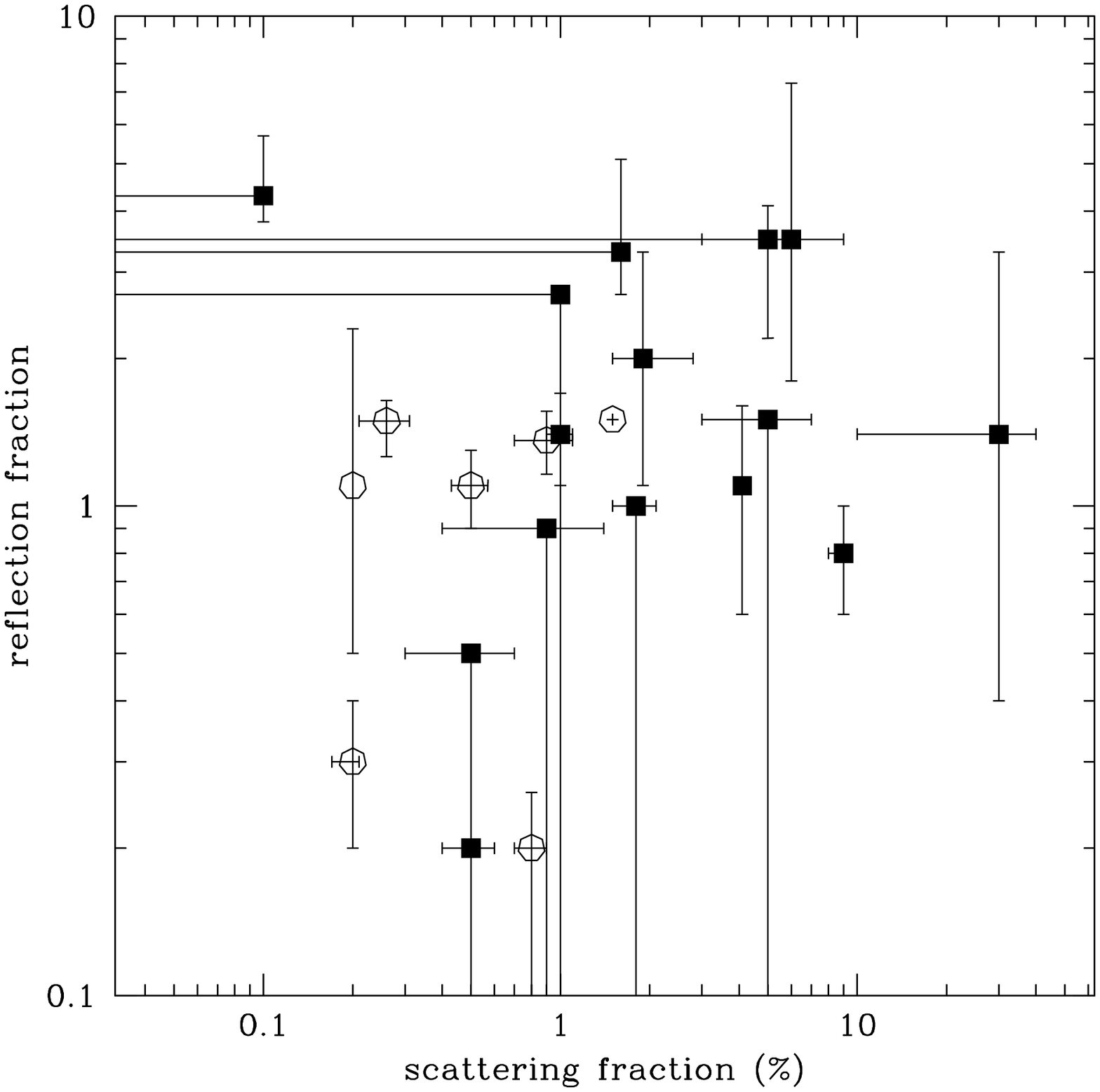}
\caption{\textit{Left panel}: Fe K$\alpha$ line EW as a function of the absorbing column density N$_{H}$. The dashed line  represents the EW values as obtained  from transmission through a uniform slab and the solid line represents the function EW(N$_{H}$)=EW$_{0}$exp($\sigma_{Fe} N_H$), see text for details.  \textit{Right panel}: Reflection fraction R as a function  of the scattering fraction f$_{scatt}$, filled boxes represent IGR sources, open circles are AGN already known.}
\label{refl_fig}
\end{figure}

\subsection{The Fe K line}
 \label{feline}
With the sole exception of IGR J17513-2011, a neutral Fe line is detected in all the sources of the sample (marginally significant for IGR J00040+7020, 95\%).
The value of the equivalent width (EW), calculated with respect to the total direct+reflected continuum\footnote{We calculated the EW as the ratio between  the integrated photon flux of the Fe line and total (direct+reflected continuum flux at the energy of the line).}, ranges between 50-2000 eV (or 500 eV excluding the Compton thick sources). The average EW value considering all sources is 430 eV (with a somewhat high standard deviation of 564 eV),  while, excluding the Compton thick objects, the average value  becomes 196 eV ($\sigma$=176 eV). 

In the left panel of Fig. \ref{refl_fig} we show the Fe K$\alpha$ EW  as a function of the absorbing column density. 
Below N$_{H}$=10$^{23.5}$cm$^{-2}$ the EW trend is almost flat, with a best fit constant value of EW$_{0}$=120$\pm$16 eV; this value is in a very good agreement with that found in other X-ray sample of Compton thin  type 2 (Fukazawa et al. 2011) as well as type 1 AGN (Bianchi et al. 2009) observed with  \suzaku, and \xmm, respectively.
However, above N$_{H}$=10$^{23.5}$cm$^{-2}$ the behaviour is different, and a clear correlation between N$_{H}$ and EW is evident, as expected from reflection from a Compton thick torus (Awaki et al. 1991). 

The question now is: what is the origin of the observed Fe K line in our sample?
The dashed line in Fig. \ref{refl_fig} represents the EW values as obtained  from transmission through a uniform slab of medium with solar abundances subtending 4$\pi$ to a continuum source of photon index $\Gamma$ = 2.0 (Leahy \& Creighton 1993).  Large EW above 10$^{23.5-24}$cm$^{-2}$ could be originated in a Compton thick torus, also responsible for the observed absorption. Since a typical torus with N$_{H}$=10$^{24}$cm$^{-2}$ should produce Fe(EW)=600-700 eV (Ghisellini et al. 1994) for solar abundances, to reproduce our values of about 1 keV, we have to consider a Compton thick medium with abundances higher than solar. On the other hand, the large EW  observed in Compton thin AGN (N$_{H}$ below 10$^{23.5}$cm$^{-2}$) are far above the dashed line, suggesting that the line is due to reflection from a material different from the absorber.
The evidence that the same Fe K$\alpha$ EW ($\sim$120 eV) is found  for Compton thin and type 1 AGN, suggests that the origin of this component is in the same medium, e.g. the BLR.
The solid line in  Fig. \ref{refl_fig}  represents the function EW(N$_{H}$)=EW$_{0}$exp($\sigma_{Fe} N_H$), with $\sigma_{Fe} $ being the photoelectric cross-section at 6.4 keV and EW$_{0}$=120 eV.
This trend well reproduces the Fe EW below 10$^{23.5}$cm$^{-2}$, and represents a scenario in which the Fe line is generated by reflection off the BLR seen along an unobscured line-of-sight, while the underlying continuum is obscured by matter with a column density $N_{H}$;  the total effect will be a larger value of the equivalent width increasing  N$_{H}$. 
In this regard, it is worth noting that in 4 sources of our sample  (IGR J07565-4139, IGR J10404-4625, IGR J12026-5349,  MCG-5-23-16) a broad Fe line is required to fit the broadband data. 
The intrinsic width of these broad lines is $\sigma_{broad}\sim$0.3 keV corresponding to a velocity of v=15000 km s$^{-1}$, that is completely compatible with an origin within the BLR.
This scenario is supported by the recent observations of rapid variation of the absorbing column density in both Compton thin and thick AGN. 
In fact, a fraction of Compton thick (thin) objects do
not intercept the torus along the line of sight,
but are caught when a Compton thick (thin) cloud located
in the BLR is passing in front of the source.
Such sources may change their status in subsequent
observations, once the thin or thick cloud has passed, explaining
some of the so-called ‘changing-look’ objects (the most clear case being NGC 1365, Risaliti et al. 2007). 

However, we stress that a Compton thin gas, also covering 4$\pi$ solid angle to the source, cannot create a Fe EW of 50-150 eV (Yaqoob etal. 2001) as observed in our sample below N$_{H}$=10$^{23.5}$cm$^{-2}$; then, a Compton thick medium like a torus (not intercepting the line of sight in Compton thin or type 1 AGN), should exist.

%

Our conclusion is that in heavily obscured Seyferts (N$_{H}$ above 10$^{23.5-24}$cm$^{-2}$), the Fe line is produced in a torus also responsible for the absorption;  for Compton thin sources (N$_{H}$ below 10$^{23.5}$cm$^{-2}$), as well as for type 1 AGN, a high fraction of the Fe K line (but not the total line photons) has to be produced in a gas  located closer to the black hole than to the torus. This region can be associated with the BLR that could also be responsible for  the observed moderate absorption.

We also investigated the X-ray Baldwin/''Iwasawa-Taniguchi'' effect (Baldwin 1977, Iwasawa \& Taniguchi, 1993) indicating the existence of an  anticorrelation between the 2--10 keV luminosity corrected for absorption and the EW of the Fe K line. This effect has been claimed in a sample of radio-quiet type 1 AGN (Bianchi et al. 2007, Page et al. 2004) in the form of EW $\varpropto$L$_{X,44}^{-0.17\pm0.03}$, with L$_{X,44}$ being the 2-10 keV luminosity in unit of 10$^{44}$ erg s$^{-1}$. This effect has not been revealed in a sample of type 2 AGN observed BeppoSAX data (Dadina 2008).
However, the physical explanation for this phenomenon is still unclear and several possibilities are generally
invoked in the literature, including a change of the ionizing continuum
and gas metallicity with luminosity (Korista et al. 1998)
or a luminosity-dependent covering factor and ionization parameter
of the BLR (Mushotzky \& Ferland 1984). On the other
hand, some studies claim that the primary physical parameter
which drives the Baldwin effect could be the accretion rate, rather than luminosity.

\begin{figure}
\centering
\includegraphics[width=0.4\linewidth]{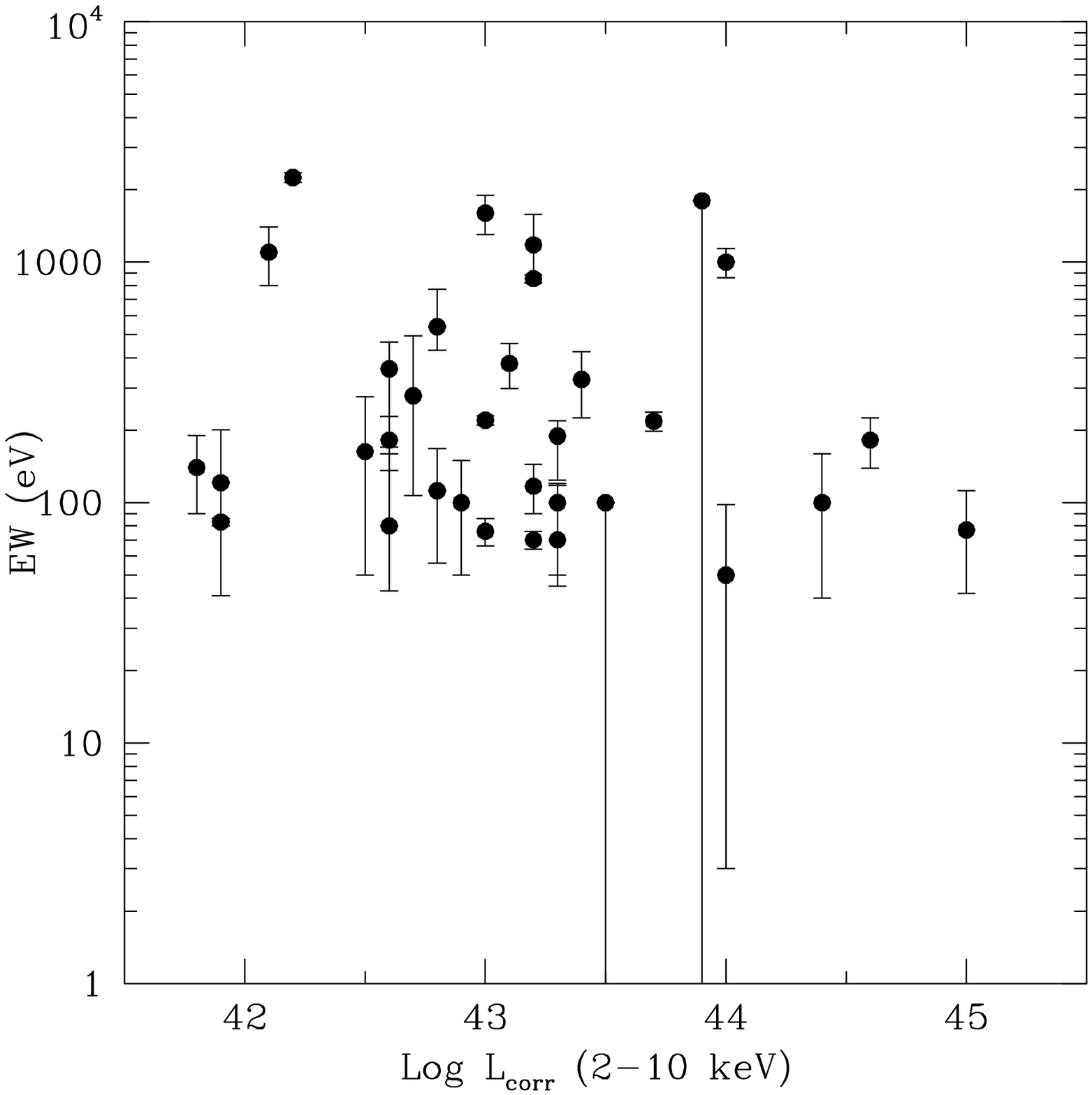}
\includegraphics[width=0.4\linewidth]{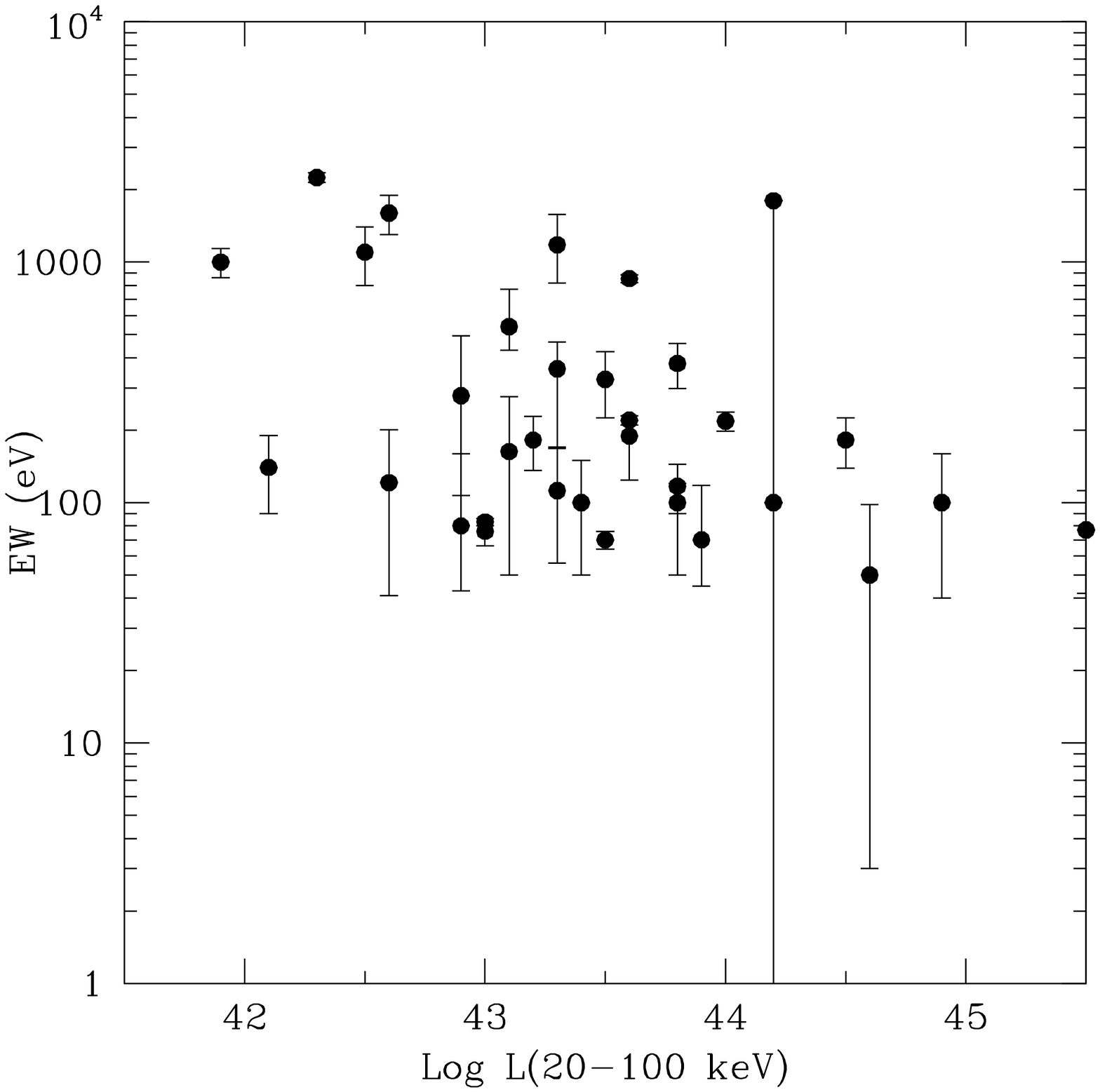}
\caption{The X-ray Baldwin effect, Iwasawa-Tanigichi effect. The EW of the Fe K line as a function  of the 2--10 keV luminosity corrected for absorption (let panel) and the 20--10 keV luminosity (right panel). }
\label{baldwin}
\end{figure}

In the left panel of Fig. \ref{baldwin} we plot the EW of the Fe line as a function of 2--10 keV luminosity corrected for absorption  (L$_{corr}$(2-10 keV)).
A  linear fit  to the data  of the form log(EW)=5.3$\pm4.8$ - (0.07$\pm$0.11)log(L$_{corr}$(2--10 keV)), indicates no evidence of a correlation (r=0.109).
The X-ray Baldwin effect is not confirmed in our sample; although, we note that using the average EW in each luminosity bin, the correlation increases to r=0.2.  
In the right panel of Fig. \ref{baldwin}, we plot the EW of the Fe line as a function of the hard X-ray luminosity in the 20--100 keV energy range. The correlation coefficient increases in this case to r=0.3, still indicating a lack of correlation.  

It is worth noting that the luminosity range we are exploring is rather narrow (log(L)=41.5-44.5), and this could make it difficult to measure any correlation as observed by Bianchi et al. (2009) over a wider luminosity range.
\subsection{Soft X-ray emission}
\label{soft X-ray}

All sources in our sample show evidence for an unabsorbed component at energies less
than 2 keV.
This soft component is likely produced
by electron scattering of the intrinsic continuum and is observed in most 
Seyfert 2 galaxies (Antonucci 1993, Matt et al. 1997). 
To parametrize the  fraction of the intrinsic flux scattered into the line of sight we used the $f_{sc}$ value. 
As shown in Table \ref{mastertab}, our results suggest a value of $f_{sc}$ of a  few per cent in all sources (with the exception of IGR J14515-5542 for which $f_{sc}$=30\% and IGR J16351-5806 with $f_{sc}$=9\%, see following discussion in this section). This value is in good agreement with that found in different samples of obscured AGN, observed with  \xmm\ and \chandra\ (Bianchi et al. 2006).

Recently, Eguchi and collaborators (2009) discussed  the evidence for an anticorrelation between the amount of relative reflection R and the scattered fraction observed below 2 keV in a small sample of obscured AGN observed with \textit{Swift}/BAT and \textit{Suzaku}.
In particular the authors claimed the existence of a new population of type 2 AGN, characterized by high R ($>$0.8) and low scattered fraction f$_{sc}$ ($< 0.5\%$) compared with the classical ones characterized by low R and high f$_{sc}$.
 We investigated this possibility within our sample and in Fig. \ref{refl_fig} (right panel) we plot the value of relative reflection R with respect to the f$_{sc}$.  
In our sample only one hard X-ray selected source, IGR J10404-4625, is characterized by high R and low f$_{sc}$;
nevertheless a linear correlation  Pearson test produces a very poor correlation coefficient (r=0.09) indicating a lack of separation between two different populations. The anticorrelation might be interpreted as due  to a degeneracy between different parameters in the fitting procedure (R and power-law normalization) as also pointed out by Comastri and collaborators (2010) who analysed a sample of obscured AGN observed with \textit{Suzaku}.

The very soft X-ray spectra of six IGR sources (namely 07565-4139, 1009-4250, 14515-5542, 16351-5806, 20286+2544, 23524+5842) clearly show more complexities than allowed by our BLM,  showing either emission lines  below 2 keV in addition to the simple scattered power--law component (see Figs. \ref{spec1}, \ref{spec2}, \ref{spec3}) or a soft power-law slope  significantly steeper than that of the hard X-ray continuum. Guainazzi \& Bianchi (2007) found that 
a photoionized plasma (probably associated with the NLR) is  an excellent description of the soft X-ray spectra of a
large sample of obscured Seyfert 2 observed with the high resolution RGS gratings on board
\xmm.  However, the available counting statistics and the limited
CCD energy resolution of our present data are not sufficient to adequately constrain the physical
properties of the emitting plasma. The short \xmm\ exposures available, also make the RGS data for these four sources useless. 
We therefore adopted a simplified approach adding a  thermal plasma component to  the   
spectra of the four sources characterised by ''soft-excess'' emission. The temperature of this medium is in the range between 0.1-1 keV, while the total integrated soft X-ray luminosity of this component ranges from 10$^{40}$ erg s$^{-1}$ to $1.4 \times 10^{41}$ erg s$^{-1}$.
This excess emission could be associated to the photoionised NLR, unresolved sources and/or diffuse emission in the host galaxy (Guainazzi et al. 2005).
\section{Type 1 $vs$ type 2 accretion properties}
\label{ty1_ty2}

\begin{table*}
\begin{flushleft}
\caption{Sub-sample of type 1+ type 2 AGN with measured Masses.}
\begin{tabular}{ccc}
\noalign{\hrule}
\noalign{\medskip}
src & Mass & reference \\
 & (M$_{\sun}$) & \\
 \hline
 type 2 & & \\
 \hline
NGC 788 & 3.2e8 & Whang \& Zhang 2007\\
NGC1068 &  10e7 & Whang \& Zhang 2007\\
NGC1142 & 2.3	e9 & Winter et al. 2009\\
MKN3 & 3.0e8 & Whang \& Zhang 2007\\
MCG-5-23-16 & 5e7 & Wandel \& Mushotzky 1986\\
NGC3281 & 4.2e8 & Winter et al. 2009\\
NGC4388 & 8.5e6 & Kuo et al. 2011\\
NGC4507 & 1.8e8 & Nicastro et al. 2003\\
NGC4945 & 1.0e6 & Greenhill et al. 1997\\
CenA & 2.0e8 & Marconi et al. 2004	\\
NGC5252 & 9.5e8 & Capetti et al. 2005\\
circinus & 1.7e6 & Marconi et al. 2004\\
NGC6300 & 2.e6 & Awaki et al. 2005\\
ESO103-G35 & 7.17e6 & Hayashida et al. 1998		\\
CygA & 2.1e9 & Marconi et al. 2004 \\
NGC7172 & 8e7& Awaki et al 2006\\
\hline
Type 1 & & \\
\hline
mcg8-11-11   & 1.5e7 & Bian \& Zhao 2003\\
MKN6 & 1.7e8 & Winter et al. 2009\\
NGC3783 & 1.1e7 & Peterson et al. 2004 \\
NGC4151 & 4.57e7 & Peterson et al. 2004 \\
NGC4593 & 9.8e6 & Peterson et al. 2004 \\
ESO511-G03 & 4.6e8 & Winter et al. 2009 \\
MCG6-30-15 & 4,5e6 & Bian \& Zhao 2003\\
NGC6814 & 1.4e8 & Winter et al. 2009 \\
IC4329A & 2.2e8 & Peterson et al. 2004 \\
4C74.26 & 1e9 & Winter et al. 2009\\
3C390 & 3.9e8 & Peterson et al. 2004\\
MR2251-178 & 5.8e8 & Brunner et al. 1997\\
IGR J07597-3842 & 2e8 & Masetti et al. 2006\\
IGR J12415-5750 & 9.2e7 & Masetti et al. 2009\\
IGR J16558-5203& 7.8e7 & Masetti et al. 2006 \\
IGR J16482-3036 & 1.4e8 & Masetti et al. 2006b\\
2E1853.7+1534  & 1.4e8 & Masetti et al. 2006c\\
\hline
\end{tabular}
\label{masstable}
\end{flushleft}

\small{}
\end{table*}

Sample selection is  the most important
issue in testing the predictions of the UM scheme and
several results inconsistent with the scheme can be explained as due
to the biases inherent in the samples (Antonucci 2002).	
Several improved and well defined samples are attempting to test the validity and limitations of Seyfert unification
 (Singh et al. 2001, Ricci et al. 2011; Brightman \& Nandra 2011), however,
issues related to the sample selection are still open. 
	
A broadband study of 41 type 1 AGN  extracted from the hard X-ray selected complete sample  (Malizia et al. 2009) has been presented by Molina et al. (2008).
Following the UM, and having corrected the 2--10 keV luminosity for the absorption effect  (L$^{corr}_{2-10keV}$), we would expect to find the same accretion properties (photon index, bolometric luminosity, Eddington ratio)  between type 1 and type 2 AGN belonging to a homogeneous sample. 
To investigate this issue we compared  the results obtained for our absorbed sample and the unobscured one by Molina et al. (2008).

\begin{figure}
\centering
\includegraphics[width=0.4\linewidth]{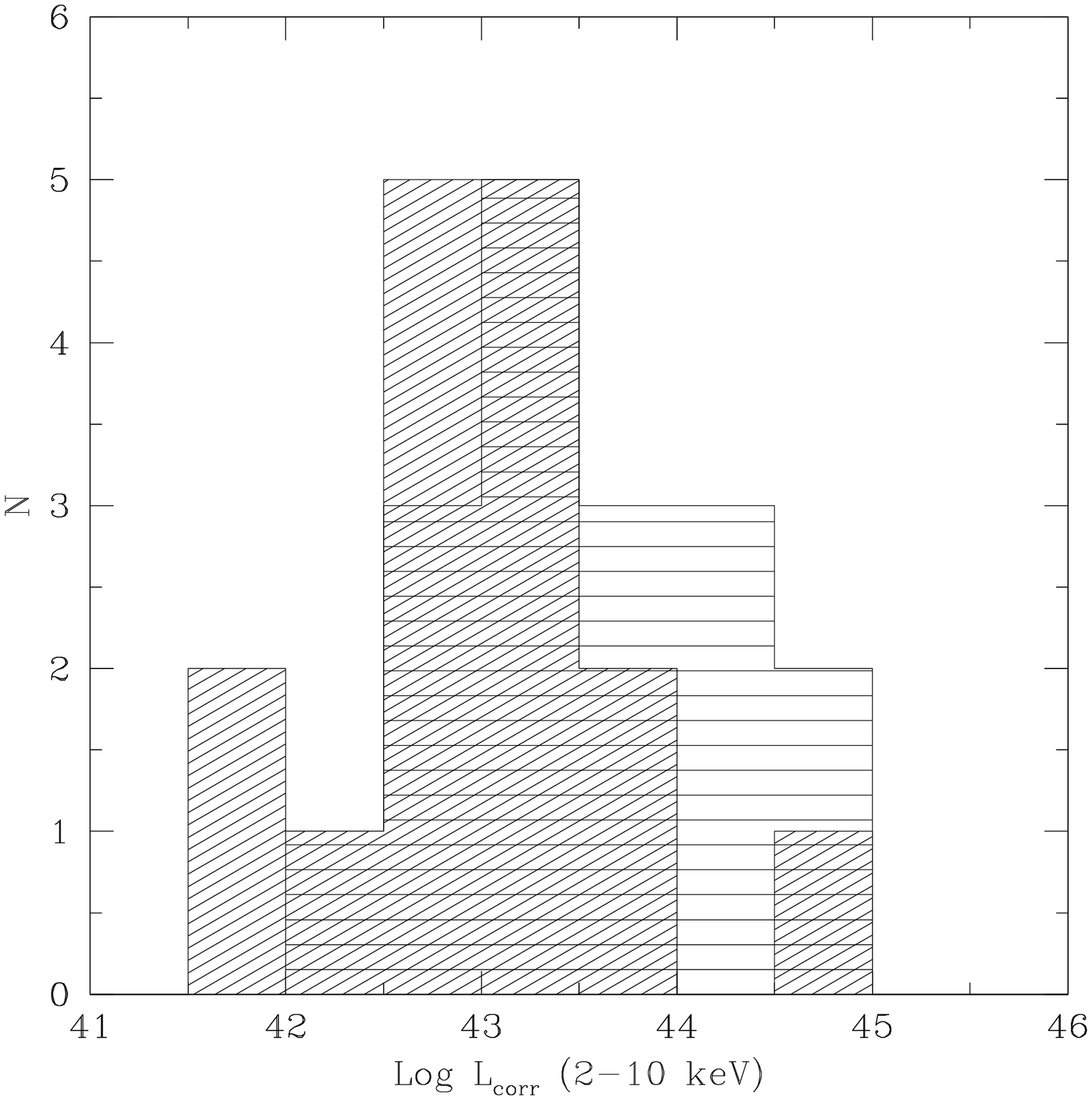}
\includegraphics[width=0.4\linewidth]{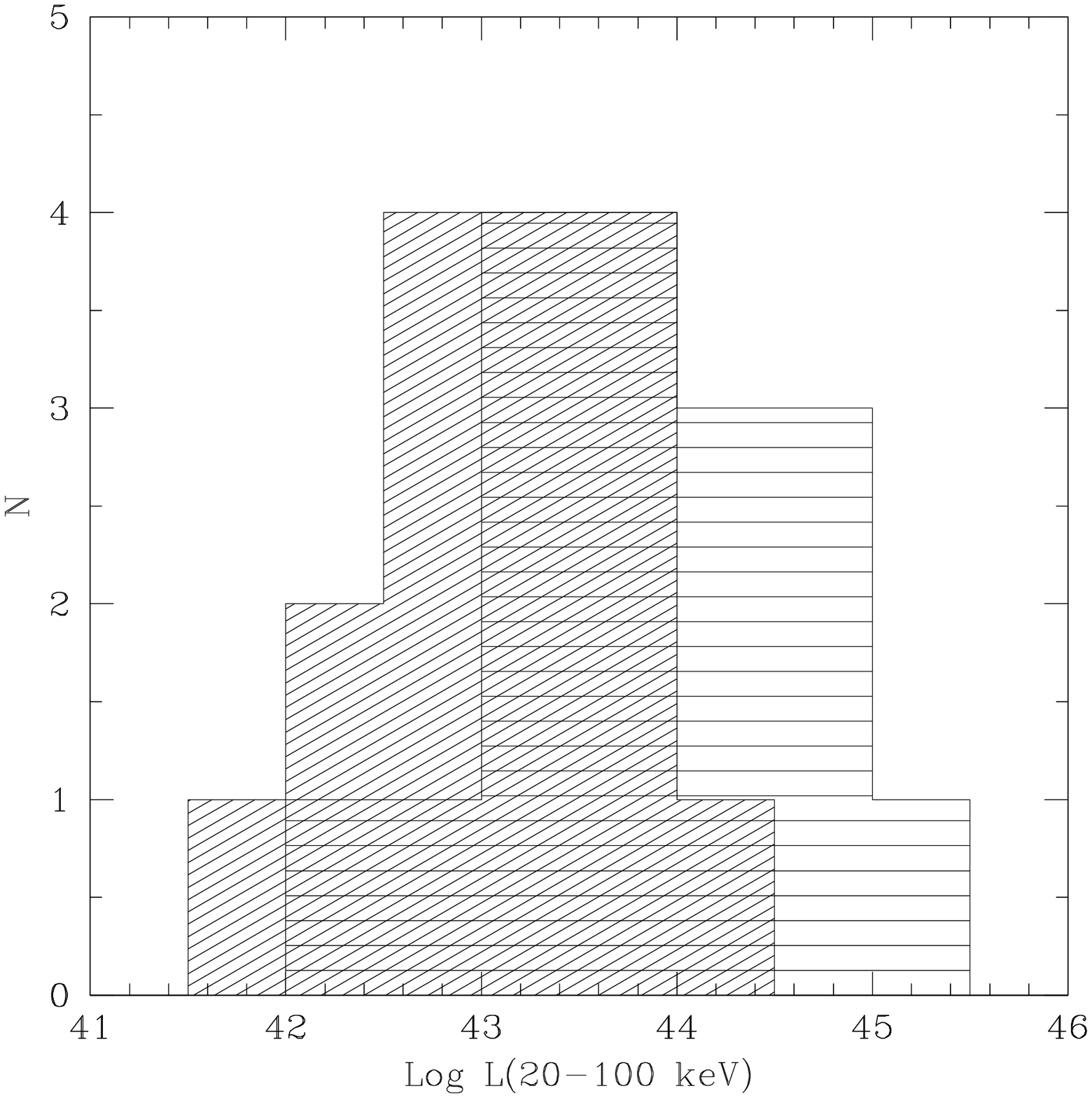}
\caption{\textit{Left panel}. Luminosity distribution (corrected for absorption) in (2--10 keV) for the sub-sample of type 1 and type 2 AGN with measured masses. Type 1 are shown as horizontal large spaced  line, type 2 are diagonal small spaced lines. \textit{Right panel}. the same distribution for the 20-100 keV luminosity L$_{20-100 keV}$. }
\label{L_distr}
\end{figure}

From the samples of obscured and unobscured AGN, 
we collected  all available published mass measurements obtained with reverberation mapping techniques, maser, X-ray variability or M-$\sigma$ relation. 
A sub-sample of 33 (17 type 1+16 type 2) AGN with known masses has been obtained, and reported in Table \ref{masstable}. In the case of unobscured AGN, Molina et al. (2008) showed that this sample can be considered as representative of the total one. 
For the obscured AGN in this work, the sub-sample we obtained is mostly representative of the known population, in fact we do not have any mass estimates for the new type 2 INTEGRAL AGN, and the sub-sample is composed by all the known sources in our original sample (excluding LEDA170194 and IC4518A). However, we have previously shown (see Sect. \ref{continuum}, \ref{reflection} and \ref{soft X-ray}) that IGR and the known objects are characterized by the same intrinsic properties. Therefore we are confident that the  sub-sample can be representative of the whole one.
 We used this hard X-ray selected sample to draw  a statistical comparison between type 1 and type 2 Seyfert of (1) the absorption corrected 2--10 keV and 20--100 keV luminosity, L$_{corr}$(2--10 keV) and L(20--100 keV); (2) the bolometric luminosity, L$_{bol}$; and (3) the Eddington ratio, L$_{bol}$/L$_{Edd}$, with the main goal of testing the prediction of the UM for AGN. In Fig. \ref{L_distr} we show  a comparison between the unabsorbed luminosities in 2--10 keV (left panel) and 20--100 keV (right panel) energy band. Horizontal large spaced lines and diagonal small spaced lines represent type 1 and type 2 class, respectively.
The distribution of the 2--10 keV luminosity has averaged value (standard deviations) of 43.52 (0.71) and 43.06 (0.72)  for type 1 and type 2, respectively. Using the  Kolmogorov-Smirnov test we do not find a significant difference in the distributions, having a probability  P=0.111 to be drawn from the same parent population.
This result is also confirmed in the higher, unbiased for absorption, energy range (20--100 keV)  with  averaged values (standard deviations) that are 43.82 (0.71) and 43.17 (0.70)  for type 1 and type 2 respectively, with a KS test probability of 0.137.
To be conservative and fully convinced to have properly corrected  the 2--10 keV luminosities for absorption, we excluded the  Compton thick sources from this analysis and, even in this case,  the KS test probability  confirms the similarity of the two populations (P=0.165).

\begin{figure}
\centering
\includegraphics[width=0.4\linewidth]{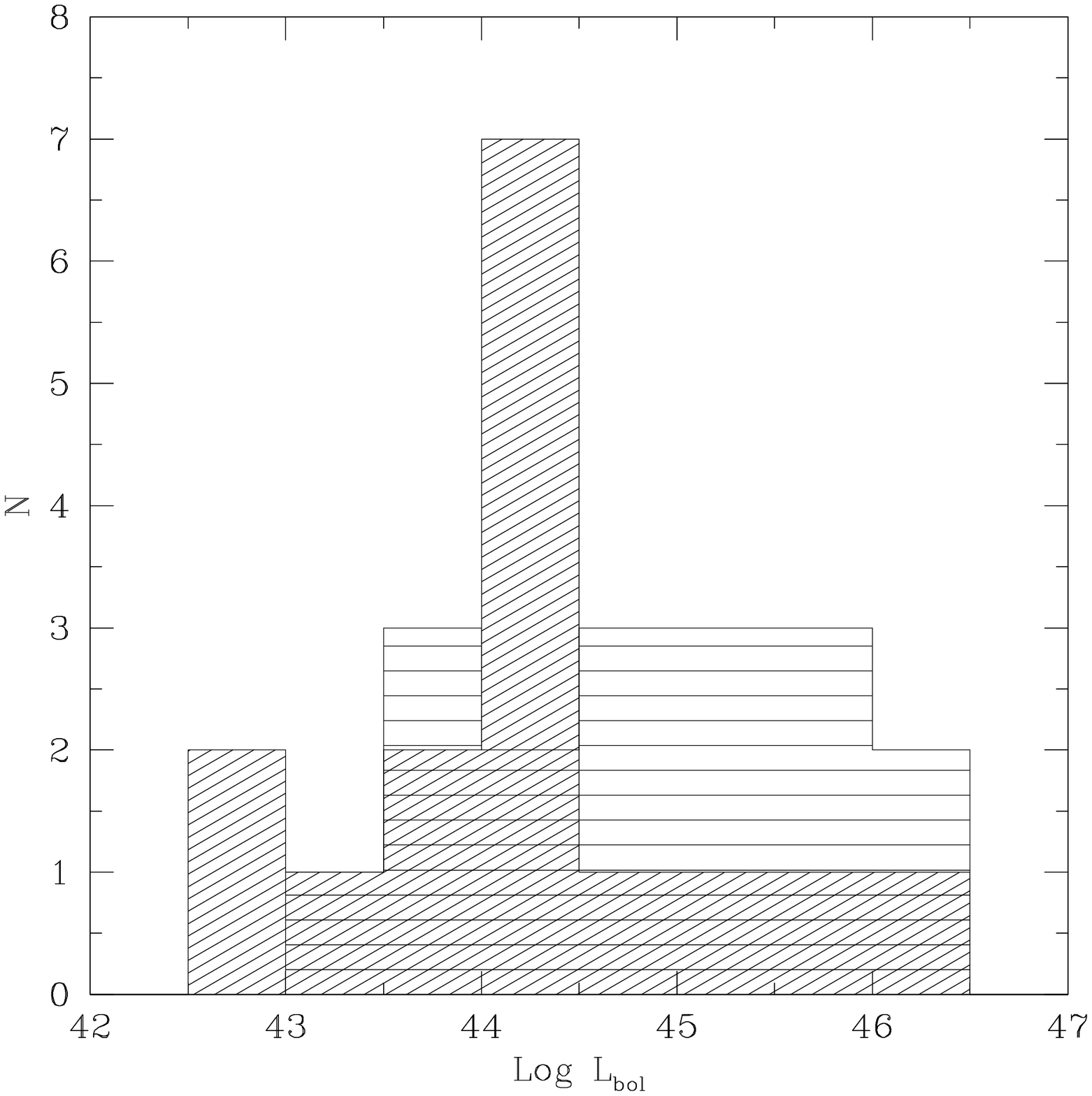}
\includegraphics[width=0.4\linewidth]{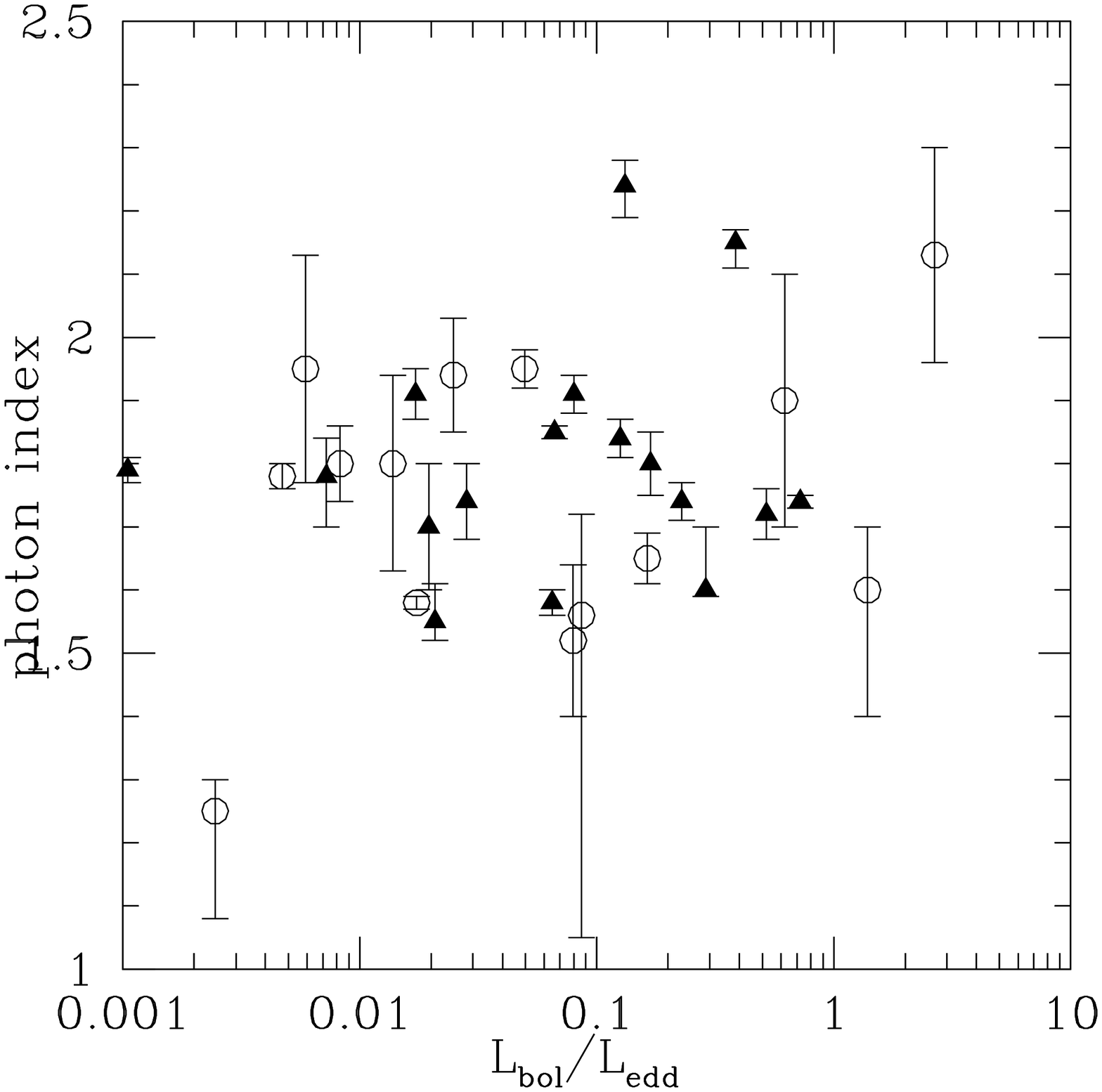}
\caption {\textit{Left panel}: Bolometric luminosity distribution for the sub-sample of type 1 and type 2 AGN with measured masses. Type 1 are shown as horizontal large spaced  line, type 2 are diagonal small spaced lines. \textit{Right panel}: Photon index as a function of the ratio L$_{bol}$/L$_{edd}$  for the sub-sample of type 1 and type 2 AGN with measured masses. Open triangles are type 1 while filled circles represent type 2 AGN.}
\label{bol}
\end{figure}

For all sources in the sub-sample we calculated the bolometric luminosity  derived from the 2--10 keV luminosities, adopting the luminosity-dependent bolometric correction given in  Marconi et al. (2004). 
In Fig. \ref{bol}, left panel,we show the distribution of L$_{bol}$ for type 1 and type 2 objects (same symbols used in Fig. \ref{L_distr}). The L$_{bol}$ average value (standrad deviation) is 44.91 (0.91) and 44.33 (0.90) for type 1 and type 2, respectively. A KS test gives a probability P=0.111 that the distributions belong to the same parent population, suggesting, even in this case, the similarity between the two classes of AGN.
In the right panel of Fig. \ref{bol} we plot the photon index of the primary power--law versus the Eddington ratio for our sub-sample.
The average value of the ratio L$_{bol}$/L$_{edd}$ (with its standard deviation) is 0.17 (0.20)  and 0.32 (0.72) for type 1 and type 2 AGN, respectively. Although the dispersion is huge, we note that, as for the luminosity, no evidence of separation in the accretion rate distribution is evident between absorbed and unabsorbed AGN (KP probability P=0.292).
These  results concerning luminosities is also reflected in the spectral properties. In fact a comparison between the primary photon index in type 1 and type 2 class does not show any evidence for a separation (see right panel in Figure \ref{bol}) being the average value of $\Gamma$ equal to 1.80 ($\sigma$=0.18) and 1.75 ($\sigma$=0.26) for type 1 and type 2 respectively. A KS test  in this case gives a probability P=0.862.

These same conclusions have  recently been reached by Singh and collaborators (2011) through the X-ray analysis of an optical selected sample of type 1 and type 2 Seyfert. 
The value of the L$_{bol}$/L$_{edd}$ ratio in our sub-sample of type 1 and type 2 AGN  range in 0.001 - 3, extending to higher values with respect to a optical selected sample (Singh et al. 2011).

We are aware that this result could be affected by two issues.
The first one concerns  the mass estimation and its error, that can be as large as 10-20 per cent. It is still debated how accurate the [OIII] FWHM is as a proxy of the stellar dispersion $\sigma$ and, especially for absorbed AGN, there are several inconsistencies between the different techniques used to measure the BH mass.
For the second caveat we assumed equal bolometric correction for both type 1 and type 2 objects, however a different bolometric correction for absorbed and unabsorbed AGN (Barger et al.\ 2005) might affect our conclusions on the bolometric luminosity distributions.
 
Finally, we also stress that, using the whole hard X-ray selected sample of  41 type 1 + 33 type 2 from Molina et al. (2008) and the present work, we find that L$_{corr}$(2--10 keV) and L(20--100 keV) have a  small  probability  (P=0.05 and 0.02, respectively) to be drawn from the same parent population. This effect is likely to be due to a bias in the sample selection; in fact,  due to the less effective area that limits the sensitivity to only bright sources, samples of Seyfert AGN based on \integral, and \textit{Swift}/BAT surveys have shown to contain a relatively large number of high luminosity and less absorbed Seyferts (Tueller et al. 2008; Treister et al. 2009; Beckmann et al. 2009).
In fact, taking into account our total type 1 + type 2 (41+33) sample, we find that the average value of the redshift is 0.05 and 0.03 for type 1 and type 2 respectively, with type 1 extending to higher redshift (higher luminosity) than type 2.

Based on the model (Nicastro 2000) in which the BLRs are formed by accretion
disc instabilities occurring in proximity to the critical radius
at which the disk changes from gas pressure dominated to
radiation pressure dominated, Nicastro and collaborators (2003)  suggested that the
absence or presence of hidden BLRs (HBLRs) is regulated by the ratio between
the X-ray luminosity and the Eddington luminosity, which, in
the accretion power scenario, is a measure of the rate at which
matter falls onto the central supermassive black hole.
Following this model, a natural transition between type 1 and type 2 AGN should then occur, depending on the accretion rate.
 According to Nicastro (2003), a value of the Eddington ratio  of 0.001, i.e. the lowest value measured in our sub-sample of type 1+type 2 AGN (see Fig. \ref{bol}, right panel), represents the threshold value above which we expect to find HBLR (for obscured objects) or BLR (for the unobscured ones).
Optical observations confirm this trend in type 1 AGN, while spectropolarimetric observations in a fraction of our sub-sample of type 2 AGN has to be performed to confirm the proposed scenario.

\begin{figure}
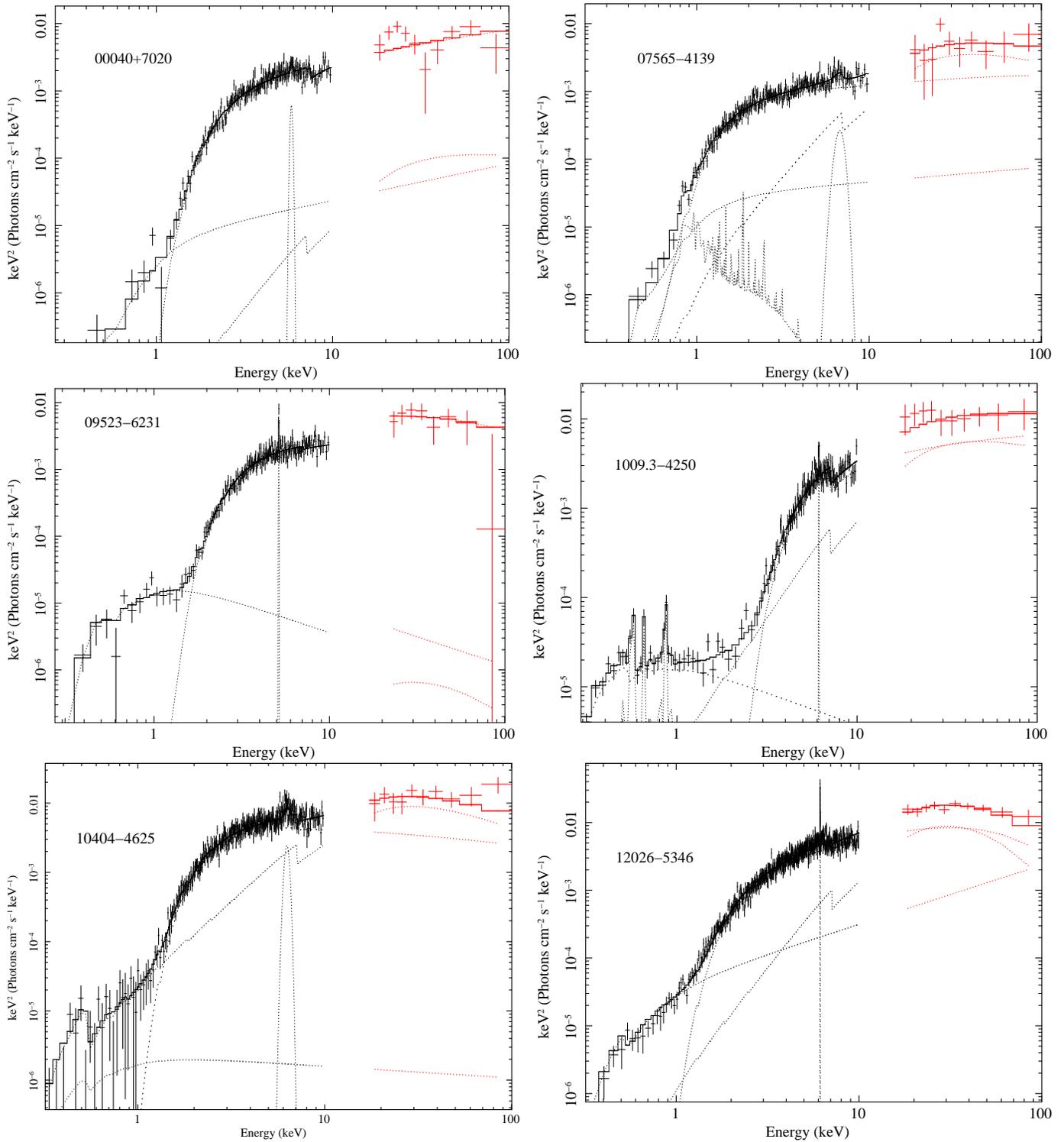

\includegraphics[width=0.4\linewidth,angle=-90]{00040b.ps}
\includegraphics[width=0.4\linewidth,angle=-90]{07565b.ps}
\includegraphics[width=0.4\linewidth,angle=-90]{09523.ps}
\includegraphics[width=0.4\linewidth,angle=-90]{1009b.ps}
\includegraphics[width=0.4\linewidth,angle=-90]{10404b.ps}
\includegraphics[width=0.4\linewidth,angle=-90]{12026b.ps}
\caption{Broadband spectra of the IGR sources in our sample.}
\label{spec1}
\end{figure}
\begin{figure}
\includegraphics[width=0.4\linewidth,angle=-90]{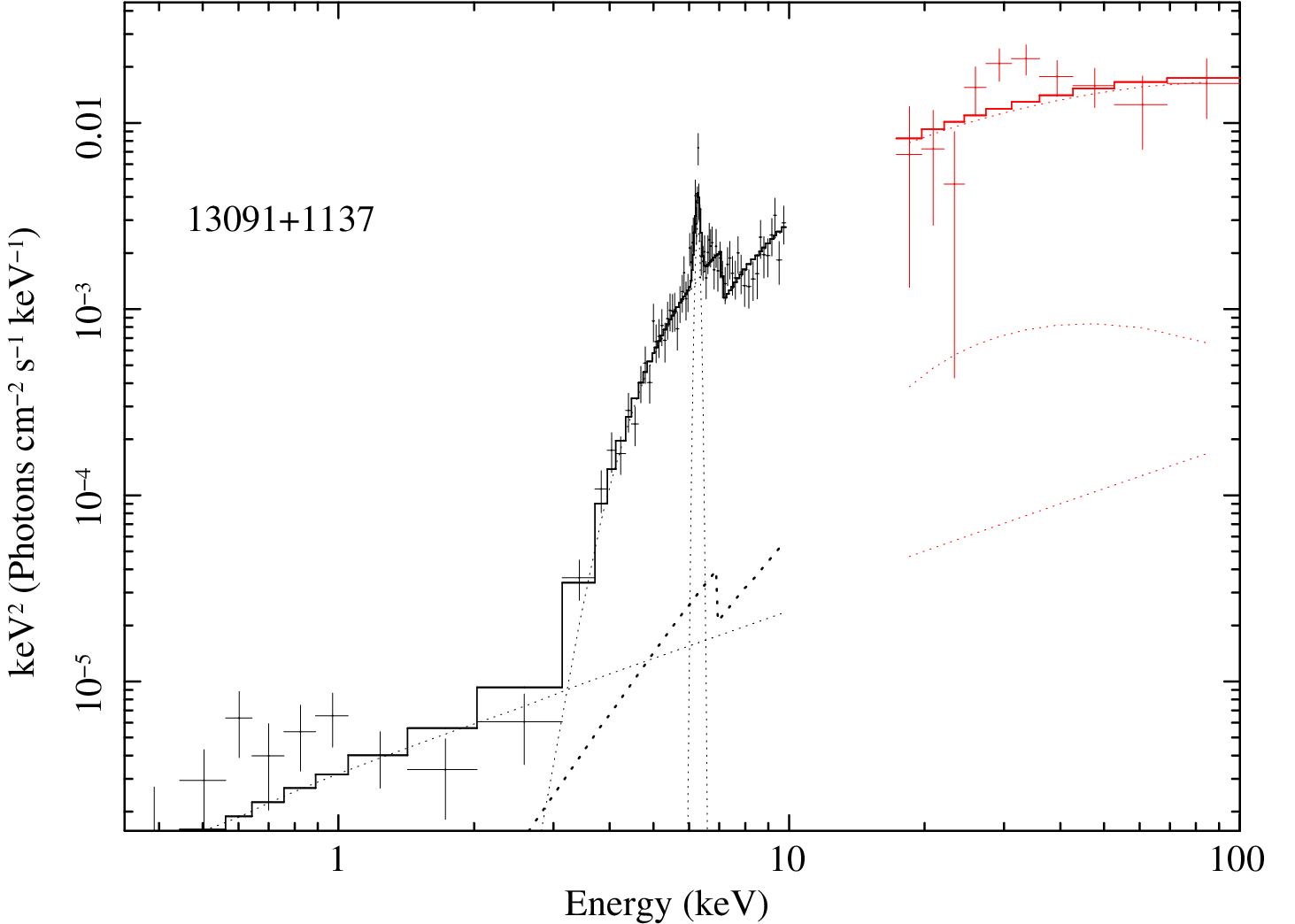}
\includegraphics[width=0.4\linewidth,angle=-90]{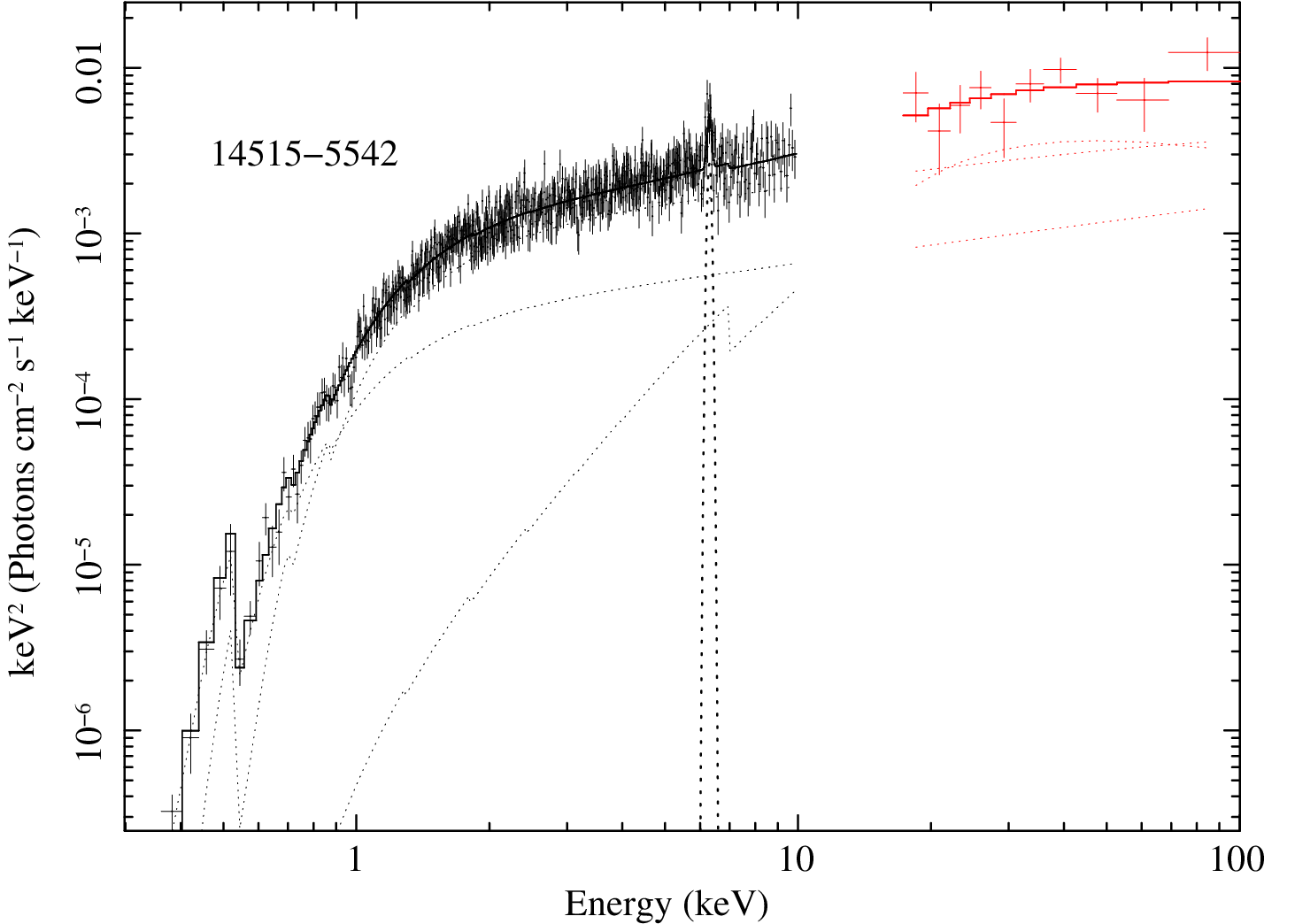}
\includegraphics[width=0.4\linewidth,angle=-90]{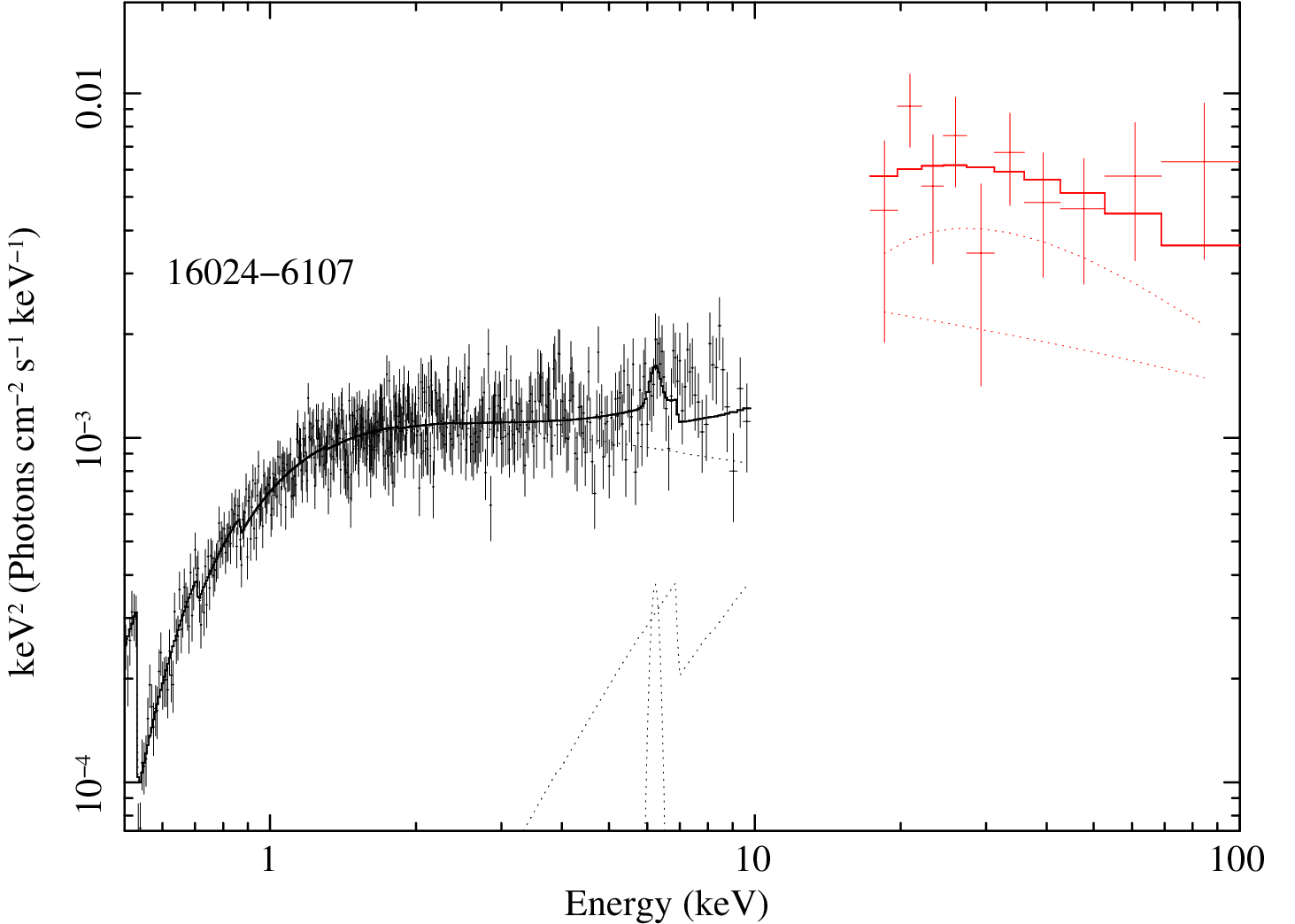}
\includegraphics[width=0.4\linewidth,angle=-90]{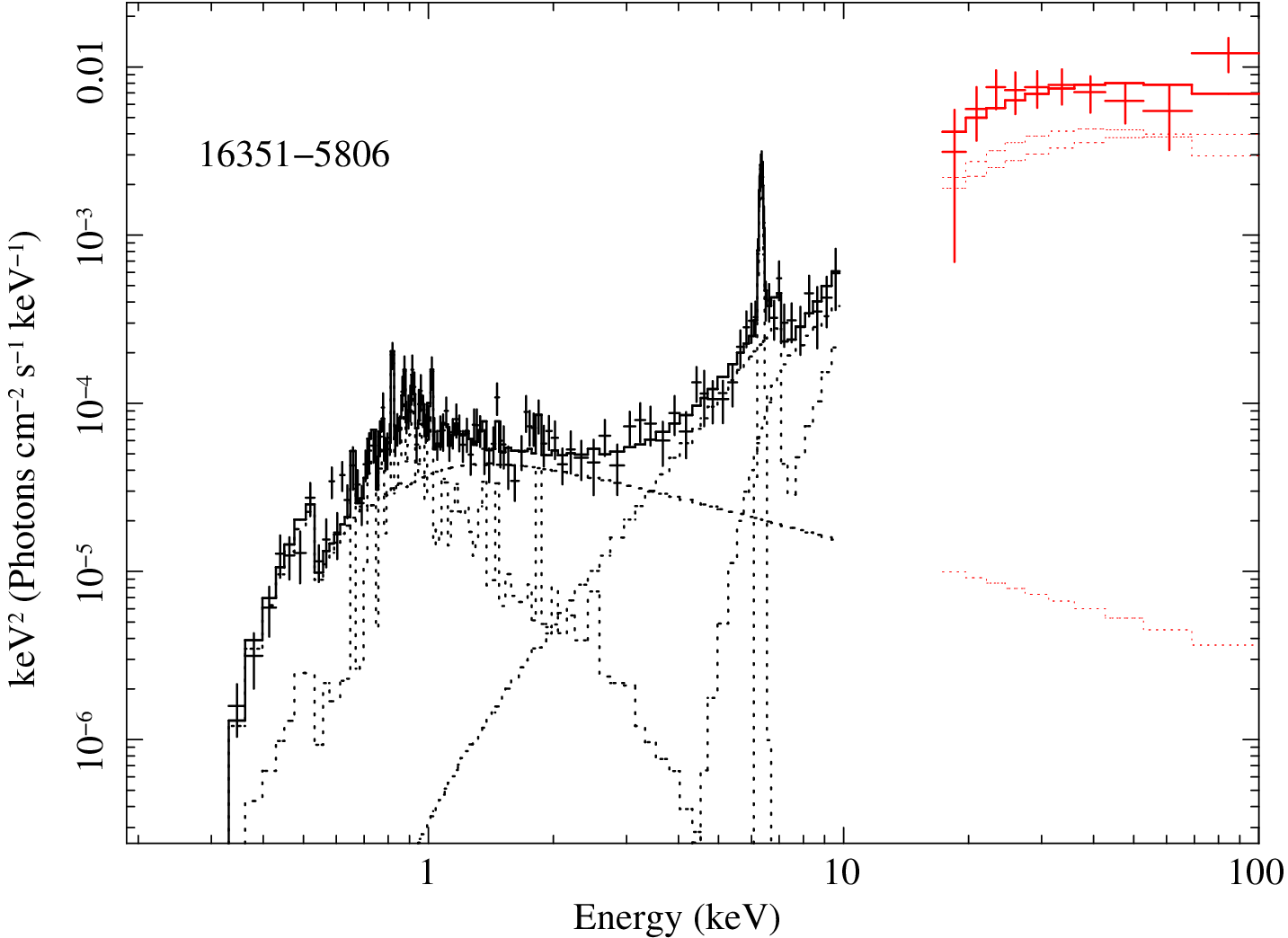}
\includegraphics[width=0.4\linewidth,angle=-90]{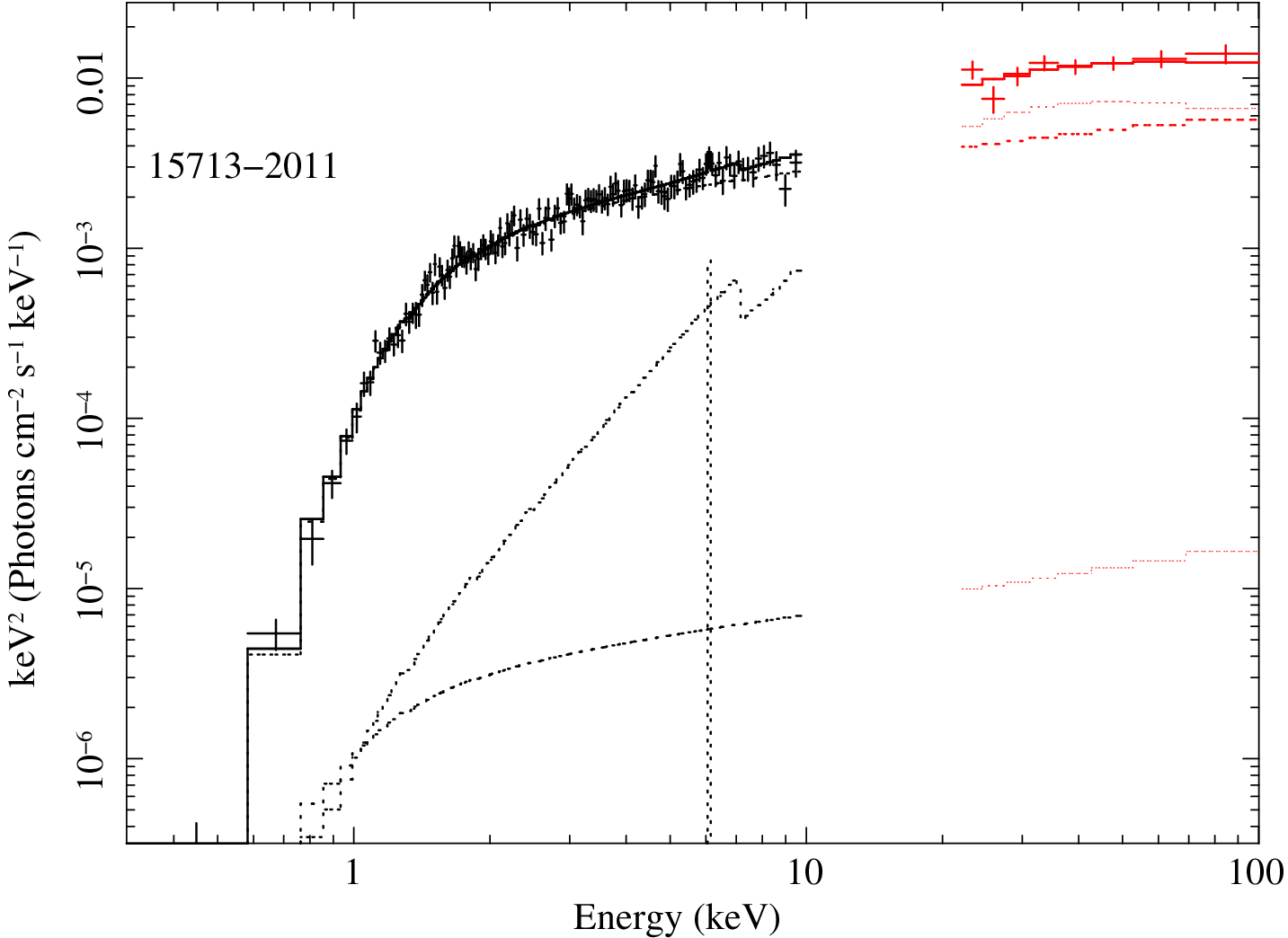}
\includegraphics[width=0.4\linewidth,angle=-90]{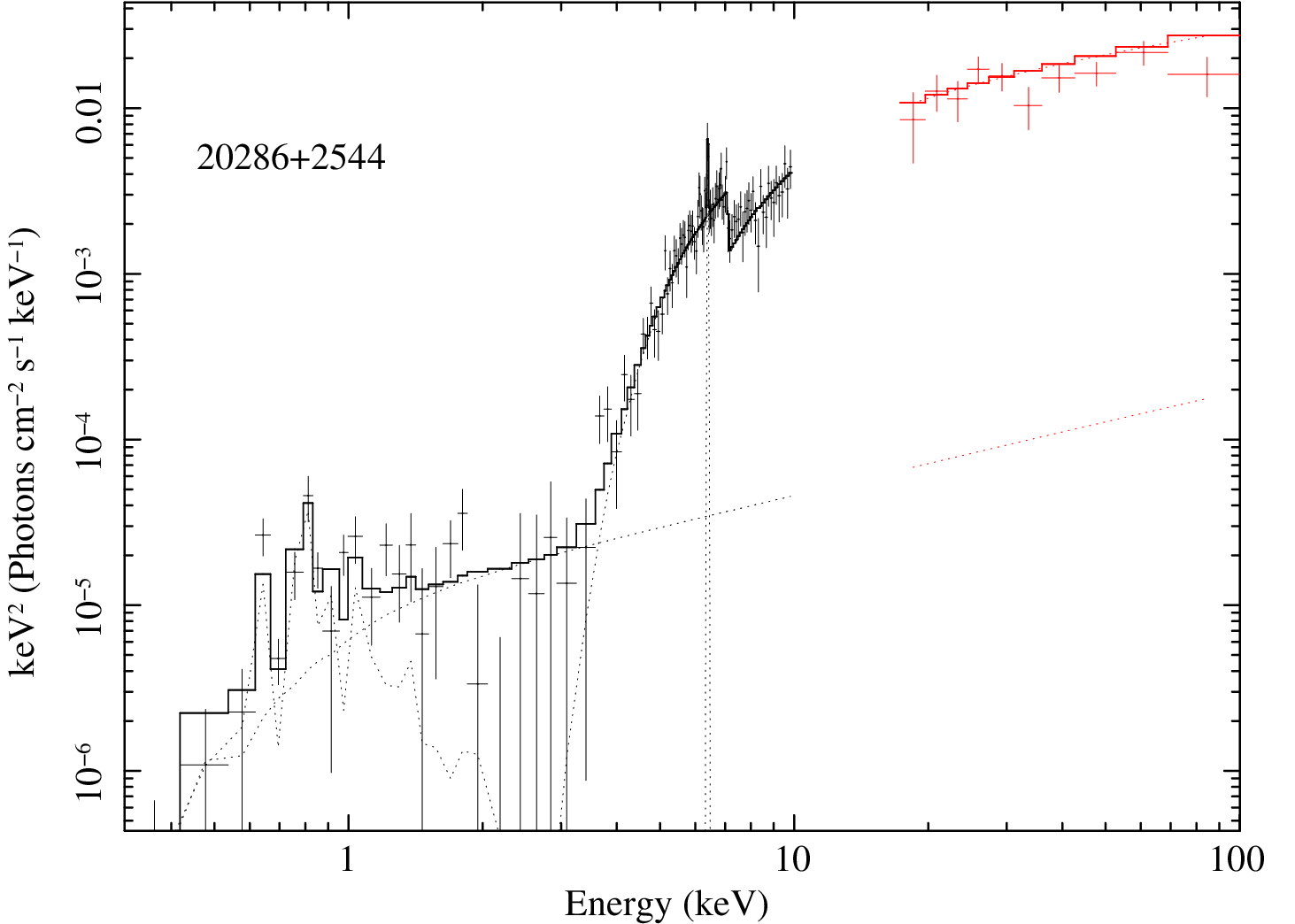}
\caption{\textit{Continued}}
\label{spec2}
\end{figure}
\begin{figure}
\includegraphics[width=0.4\linewidth,angle=-90]{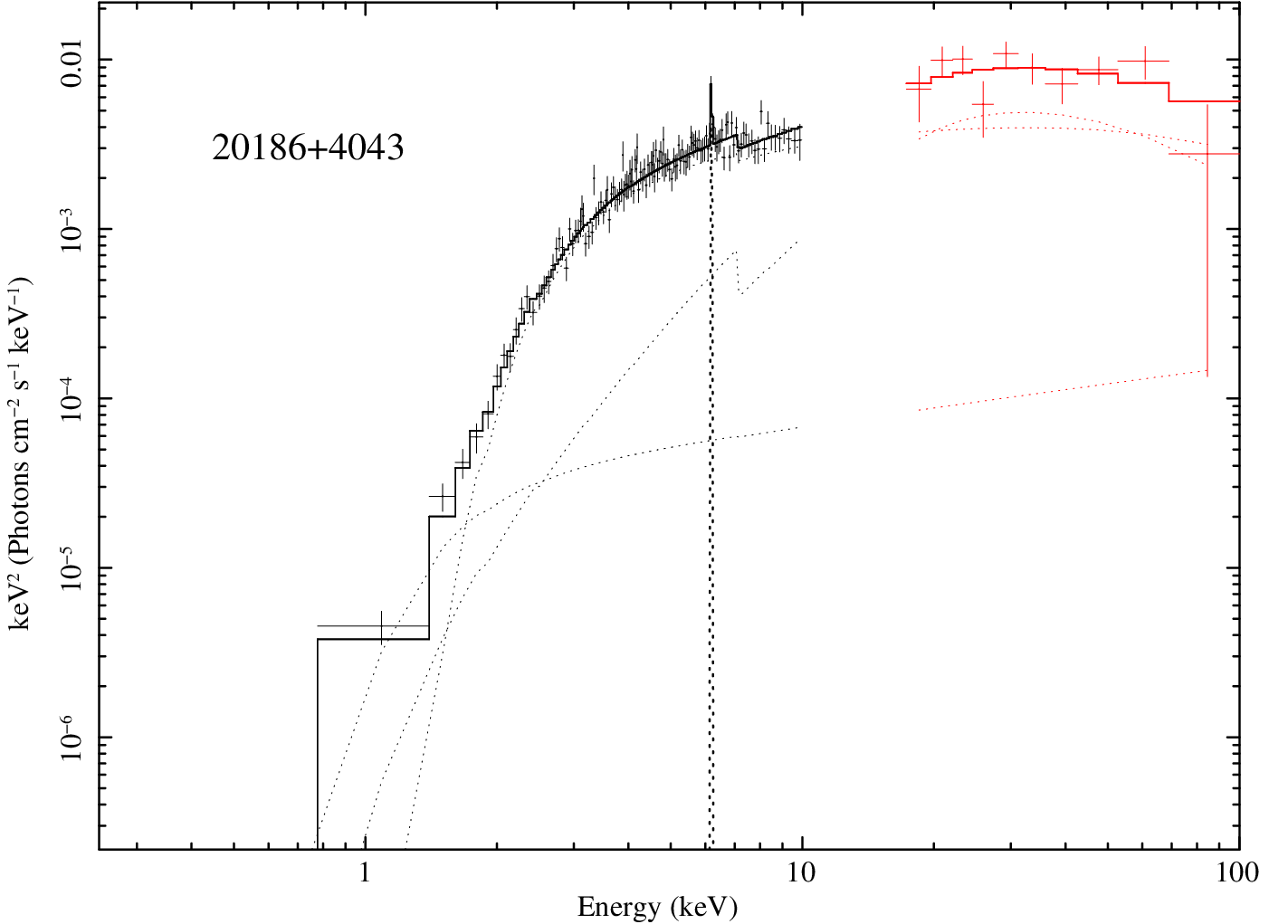}
\includegraphics[width=0.4\linewidth,angle=-90]{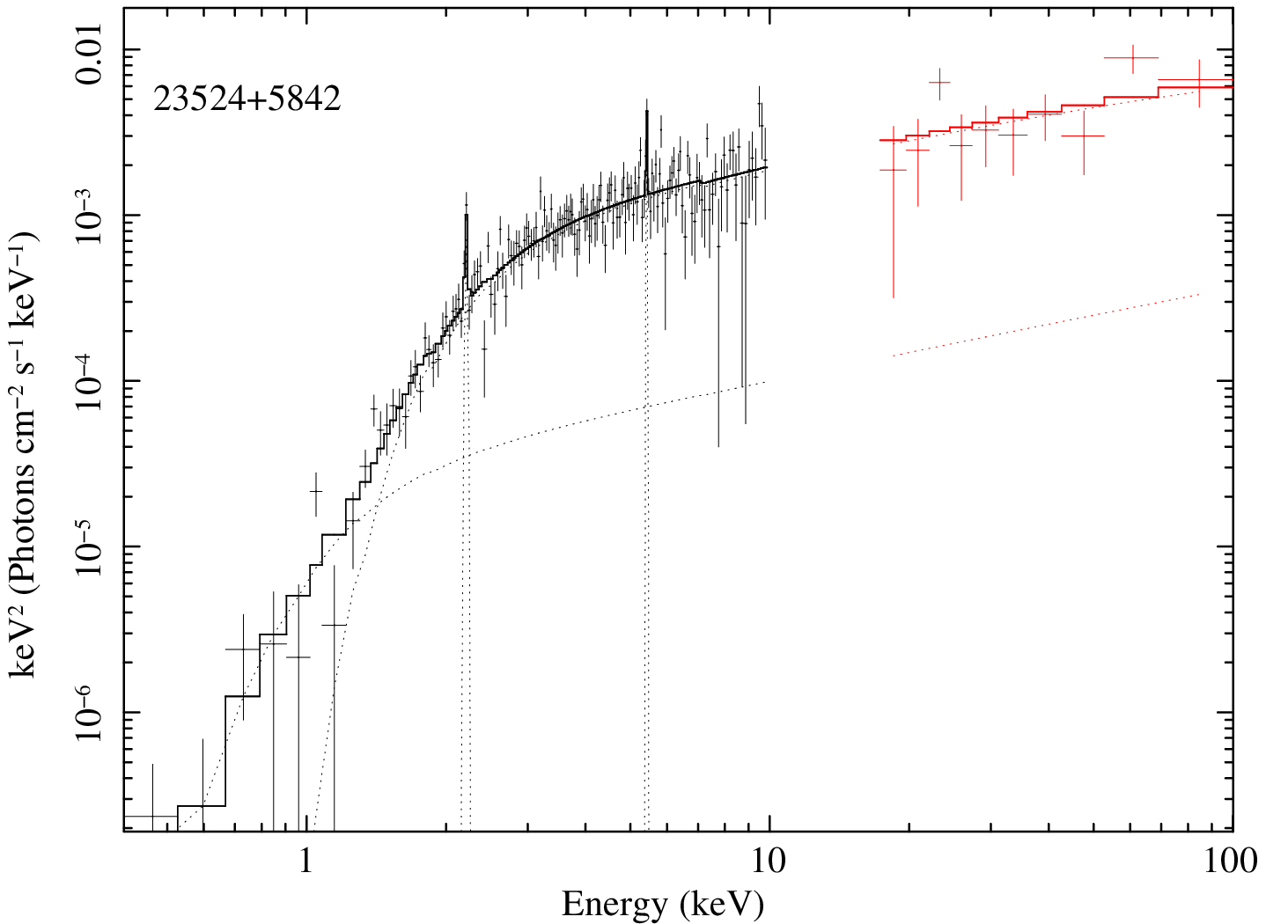}
\includegraphics[width=0.4\linewidth,angle=-90]{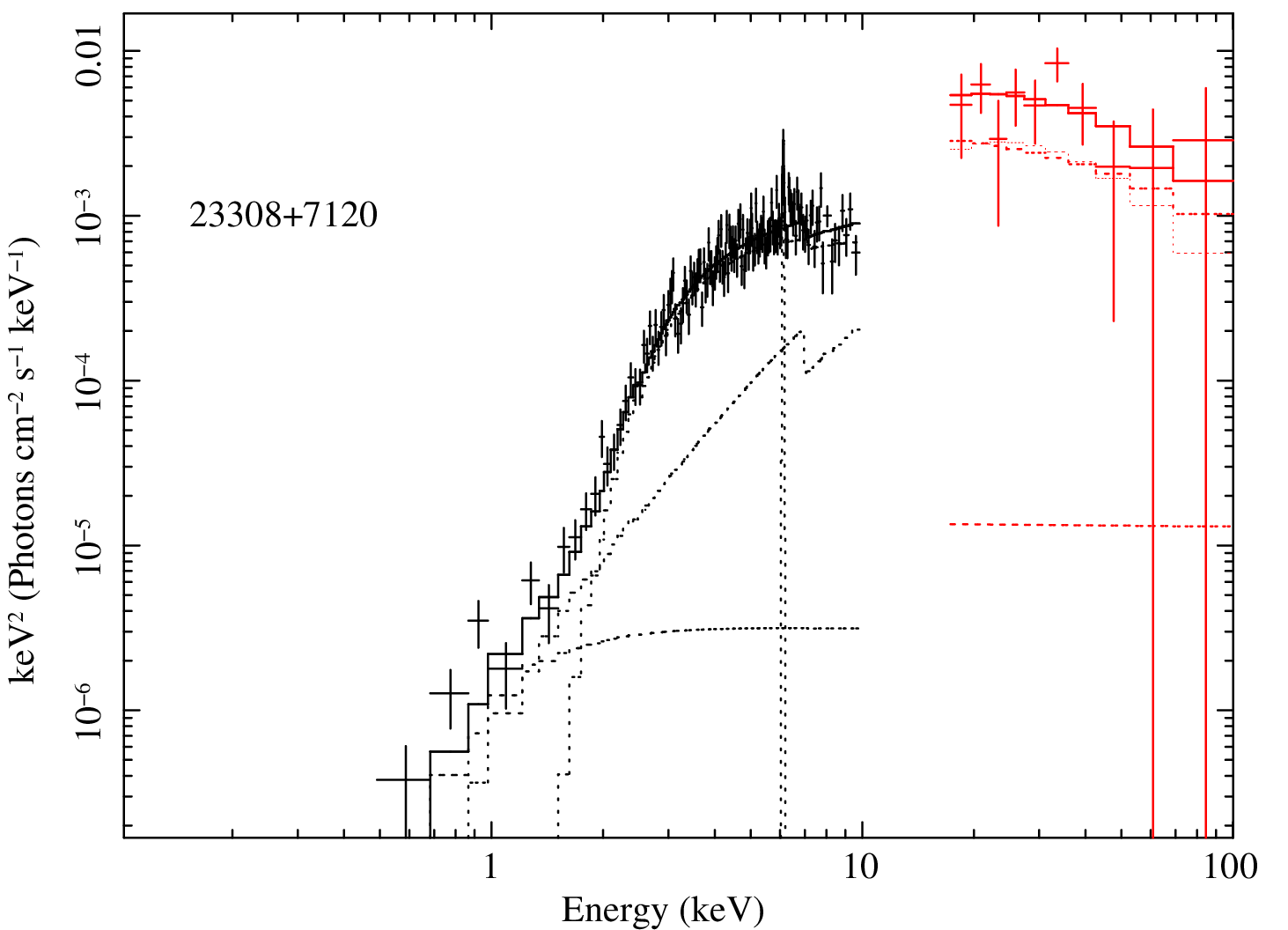}
\caption{\textit{Continued}}
\label{spec3}
\end{figure}

\section{Conclusion}
\label{conclusion}

In this paper we have discussed the broadband X-ray properties of a sample of 33 absorbed hard X-ray selected AGNs.
We selected a baseline model to reproduce the 0.3--100 keV data consisting of
 an absorbed primary emission with a high energy cutoff and its scattered fraction below 2-3 keV, plus the Compton reflection features (Compton hump and Fe line emission).
The broadband data coupled with the good energy resolution at the iron line energy range, allowed us to get a very good characterization of the primary continuum and of all the spectral features.

The main results of this study can be summarized as follows:

\begin{itemize}

\item 
We measure a high-energy rollover below 150 keV  in 30\% of the sample (10/33) and estimate a lower limit for the other sources is well below 300 keV. Our analysis suggests that the high energy cut-off is ubiquitous in the X-ray spectra of type 2 AGN.
This evidence  puts strong constraints on XRB synthesis   model.

\item
The hard X-ray selection allowed us to investigate the N$_{H}$ distribution in a sample much less severely biased by  absorption with respect to those selected in 2--10 keV band.  When compared with 2-10 keV selected samples, our objects are characterized  by larger column density; this evidence confirms the capability of hard X-ray observations for selecting obscured objects. In addition, the high S/N ratio of our wide band data  allowed us to isolate the Compton thick sources in a diagnostic plot (softness ratio F$_{oss}$(2--10 keV)/F(20--100 keV) $vs$ N$_{H}$). 

\item
The value of the Fe K$\alpha$ EW as a function of the absorbing column density and the broad profile of the line in 4 sources, support the fact that (at least) a fraction  of the Fe line has to be produced in a Compton thin gas, likely associated to the BLR. This gas intercepts the line of sight and should also be responsible for the moderate absorption observed in the mildly absorbed sources 
($\sim$10$^{23}$ cm$^{-2}$).
It is worth noting that we still need a Compton thick torus (not intercepting the line of sight) in moderately absorbed AGN to (1) produce the reflection continuum observed with \integral\ above 10 keV and (2) to contribute to the production of  the Fe line emission.
To confirm this scenario, broadband time-resolved variability study are needed, to directly observe the N$_{H}$ and Fe EW variations. These observational campaigns have been performed, so far,  in just a few sources like NGC1365, NGC7582, i.e.  the so called ''changing look'' AGN (Risaliti et al. 2009 and reference therein, Bianchi et al. 2009).

\item
A comparison with a hard X-ray selected sample of type 1 AGN shows that obscured and unobscured Seyfert are characterized by the same nuclear and accretion properties, i.e. luminosity (this result holds in both 2--10 keV and 20--100 keV bands),
 bolometic luminosity,  Eddington ratio and primary photon index.  This evidence suggests that type 1 and type 2 objects should share the same central core.

\item
The Eddington ratio would suggest, according to the model of Nicastro (2000), that all our type 2 AGN hide a BLR. 
Spectropolarimetric measurements are necessary to definitely establish the presence of HBLR clouds in the INTEGRAL AGN complete sample.

\end{itemize}

More accurate broadband data up to 80 keV will became available in the near future with \textit{NuStar} (Harris et al. 2010) to better investigate these issues.

\section*{Acknowledgements}
We thank the anonymous referee for her/his  suggestions which helped improving the paper. ADR acknowledges L. Piro for useful discussions about Fe lines.
The italian authors acknowledge the ASI financial support via grant ASI-INAF I/033/10/0 and I/009/10/0.

{}
\end{document}